\documentclass[a4paper,UKenglish,cleveref, autoref, thm-restate]{lipics-v2021}

\pdfoutput=1 
\hideLIPIcs  

\usepackage{float}

\usepackage{amsmath, mathtools, amssymb}
\usepackage{algorithm, times, yhmath}
\usepackage[noend]{algpseudocode}
\usepackage{nicefrac}
\usepackage[many]{tcolorbox}

\usepackage[dvipsnames,table]{xcolor}

\newtcolorbox{blackbox}[1][]{
    colframe=black,
    colback=lightgray!10,
    coltext=black,
    breakable,
    #1
}

\usepackage{tikz}

\usetikzlibrary{calc, shapes.geometric, arrows, automata, decorations.pathreplacing, arrows.meta}

\usepackage{hyperref}
\hypersetup{
    colorlinks=true,
    linkcolor=blue,
    filecolor=blue,      
    urlcolor=blue,
    citecolor=blue,
    pdftitle={Improved Bounds for Estimating Average Degree of Graphs},
    pdfpagemode=FullScreen,
    }
\usepackage[capitalise]{cleveref}
\usepackage[textwidth=4em,]{todonotes}

\newcommand{\sA}{\mathcal{A}}

\newcommand{\sD}{\mathcal{D}}

\newcommand{\heavyverts}{H}

\newcommand{\heavyvert}{h}

\newcommand{\lightverts}{L}
\newcommand{\lightvert}{l}
\newcommand{\heavyedges}{\edgeset(\heavyverts)}
\newcommand{\lightedges}{\edgeset(\lightverts)}
\newcommand{\hledges}{\edgeset(\heavyverts,\lightverts)}
\newcommand{\hedgecnt}{\edgecount_{\heavyverts}}
\newcommand{\ledgecnt}{\edgecount_{\lightverts}}

\newcommand{\dlest}{\widehat{w}_\lightverts}
\newcommand{\dltrue}{w_\lightverts}
\newcommand{\ldens}{\rho_{\lightverts}}
\newcommand{\estldens}{\widehat{\ldens}}
\newcommand{\degq}{q}
\newcommand{\coinq}{r}
\newcommand{\lbq}{q}

\newcommand{\lbclqv}{\vertexset_{c}}
\newcommand{\lbmatv}{\vertexset_{m}}
\newcommand{\lbclqe}{\edgeset_{c}}
\newcommand{\lbmate}{\edgeset_{m}}
\newcommand{\lbclqcnt}{k}
\newcommand{\lbdegq}{Q_{d}}
\newcommand{\lbreq}{Q_{e}}

\newcommand{\degest}{\widehat{\degree{}}}

\newcommand{\ldensest}{\hyperref[Alg: Light Density Estimator]{\texttt{CoinToss}}}

\newcommand{\nldensest}{\hyperref[Alg: Light Density Estimator Noisy]{\texttt{NoisyCoinToss}}}

\newcommand{\meanest}{\hyperref[Alg: Light Degree Estimator]{\texttt{MeanEst}}}

\newcommand{\alladv}{\hyperref[Algorithm: All Advice]{\texttt{AllAdvice}}}

\newcommand{\thradv}{\hyperref[Algorithm: Thr Advice]{\texttt{ThresholdAdvice}}}

\newcommand{\noadv}{\hyperref[Algorithm: No Advice]{\texttt{NoAdvice}}}

\newcommand{\degbnd}{\widetilde{\degree{}}}
\newcommand{\degub}{\Bar{\degree{}}}

\newcommand{\qcolor}[1]{\textcolor{violet}{#1}}

\newcommand{\ptp}{\texttt{PTP}}
\newcommand{\ptpprob}{k}
\newcommand{\ptpsep}{\gamma}
\newcommand{\ptpdone}{\distribution_0}
\newcommand{\ptpdtwo}{\distribution_1}
\newcommand{\defptp}{\ptp_{\stringlength,\ptpprob,\ptpsep}}

\newcommand{\emptycell}{\cgray{-}}

\newcommand{\cgray}{\cellcolor{gray!20}}

\newcommand{\nrandedgeq}{\texttt{AppRandEdge}}
\newcommand{\edgenoise}{\eta}
\newcommand{\field}[1]{\mathbb{#1}}

\newcommand{\norm}[1]{\left\|{#1}\right\|}

\newcommand{\ceil}[1]{\left\lceil{#1}\right\rceil}

\newcommand{\sequence}[2]{{#1}_1,{#1}_2,...,{#1}_{#2}}
\newcommand{\fbrac}[1]{\left({#1}\right)}
\newcommand{\sbrac}[1]{\left\{{#1}\right\}}
\newcommand{\tbrac}[1]{\left[{#1}\right]}
\newcommand{\abs}[1]{\left|{#1}\right|}
\newcommand{\size}[1]{\left|{#1}\right|}
\newcommand{\constant}{c}

\DeclareMathOperator*{\E}{\field{E}}
\DeclareMathOperator*{\Var}{\mathrm{Var}}
\DeclareMathOperator*{\Cov}{\mathrm{Cov}}

\newcommand{\distribution}{\sD}

\newcommand{\uniform}{\mathrm{Unif}}

\newcommand{\approxerror}{\varepsilon}
\newcommand{\confidence}{\delta}

\newcommand{\bigo}[1]{O\fbrac{{#1}}}
\newcommand{\bigot}[1]{\widetilde{O}\fbrac{{#1}}}
\newcommand{\bigoten}[1]{\widetilde{O}_{\approxerror,\log\vertexcount}\fbrac{{#1}}}
\newcommand{\smallo}[1]{o\fbrac{{#1}}}

\newcommand{\bigomega}[1]{\Omega\fbrac{{#1}}}

\newcommand{\bigtheta}[1]{\Theta\fbrac{{#1}}}

\newcommand{\graph}{G}
\newcommand{\vertexset}{V}
\newcommand{\edgeset}{E}
\newcommand{\vertexcount}{n}
\newcommand{\edgecount}{m}
\newcommand{\vertex}{v}
\newcommand{\altvertex}{u}
\newcommand{\edge}{e}
\newcommand{\neighbour}[1]{{\sf N}\fbrac{#1}}
\newcommand{\degree}[1]{d_{#1}}

\newcommand{\arboricity}{\alpha}

\newcommand{\randedgeq}{\texttt{RandEdge}}
\newcommand{\neighbourq}{\texttt{Neighbour}}
\newcommand{\edgeexistsq}{\texttt{Pair}}
\newcommand{\fullnbrq}{\texttt{FullNbr}}
\newcommand{\degreeq}{\texttt{Degree}}

\newcommand{\stringlength}{M}

\newcommand{\alicestring}{x}

\newcommand{\threshold}{\tau}

\newcommand{\poly}[1]{\mathrm{poly}\fbrac{#1}}
\newcommand{\polylogneps}{\poly{\approxerror^{-1}\log\vertexcount}}

\title{Improved Bounds for Estimating Average Degree of Graphs} 

\author{Debarshi Chanda}{Indian Statistical Institute, Kolkata, India \and \url{https://sites.google.com/view/debarshi-ch/}}{debarshi.chanda.1997@gmail.com}{https://orcid.org/0009-0001-2197-6366}{}

\authorrunning{D. Chanda}

\Copyright{Jane Open Access and Joan R. Public} 



\category{} 

\relatedversion{} 

\nolinenumbers

\EventEditors{John Q. Open and Joan R. Access}
\EventNoEds{2}
\EventLongTitle{42nd Conference on Very Important Topics (CVIT 2016)}
\EventShortTitle{CVIT 2016}
\EventAcronym{CVIT}
\EventYear{2016}
\EventDate{December 24--27, 2016}
\EventLocation{Little Whinging, United Kingdom}
\EventLogo{}
\SeriesVolume{42}
\ArticleNo{23}

\usepackage{wrapfig}

\begin{document}

\maketitle

\begin{abstract}
We propose a randomized parameter-oblivious algorithm that, given query access to a graph $G$ with arboricity $\alpha$, and average degree $d$, makes $\widetilde{O}\left(\nicefrac{\alpha}{\varepsilon^2d}\right)$\footnote{By $\widetilde{O}(\cdot)$, we hide only $\mathrm{poly}{\log \alpha}$ terms. In particular, we do not hide $\mathrm{poly}{\log n}$ factors.} \texttt{Degree} and $\widetilde{O}\left(\nicefrac{1}{\varepsilon^2}\right)$ \texttt{Random Edge} queries to obtain an estimate $\widehat{d}$ satisfying $\widehat{d} \in (1\pm\varepsilon)d$. This also yields an algorithm that uses $\widetilde{O}\left(\nicefrac{1}{\varepsilon^2}\min(d,\nicefrac{\alpha}{d})\right)$ \texttt{Degree} and \texttt{Random Edge} queries. This improves the $\widetilde{O}_{\varepsilon,\log n}(\nicefrac{1}{\varepsilon^2}$ $\min(d,\sqrt{\nicefrac{n}{d}}))$\footnote{The $\widetilde{O}_{\varepsilon,\log n}(\cdot)$ notation hides $\mathrm{poly}(\varepsilon^{-1}\log n)$ terms.} query algorithm of [Beretta et al., SODA 2026] that has access to \texttt{Degree}, \texttt{Neighbour}, and \texttt{Random Edge} queries.  
    
    We also show that any algorithm that has access to \texttt{Degree}, \texttt{Neighbour}, and \texttt{Random Edge} queries, must make at least $\Omega\left(\min\left(d,\nicefrac{\alpha}{d}\right)\right)$ queries, even with the knowledge of graph parameters $n$ and $\alpha$. We also show that even with \texttt{Pair} and \texttt{FullNbr} queries, an algorithm must make $\Omega\left(\min\left(d,\nicefrac{\alpha}{d}\right)\right)$ queries for certain ranges of $\alpha$. For the case when $n$ is unknown, we show that any algorithm with access to all the queries discussed above must make $\Omega\left(\min\left(d,\nicefrac{\alpha}{d}\right)\right)$ queries. 
    
    We also establish $\varepsilon$-sensitive lower bounds for the problem. We establish that any sublinear algorithm using \texttt{Degree} and \texttt{Random Edge} queries must make at least $\Omega(\nicefrac{1}{\varepsilon})$ \texttt{Random Edge} queries. Furthermore, any algorithm having access to \texttt{Degree}, \texttt{FullNbr}, and \texttt{Random Edge} queries that makes $o(\nicefrac{1}{\varepsilon^2})$ \texttt{Random Edge} queries must make $\Omega(\nicefrac{\alpha}{\varepsilon^2d})$ \texttt{Degree} and \texttt{FullNbr} queries. This settles the query complexity of estimating average degree using \texttt{Degree} and \texttt{FullNbr} queries. 
\end{abstract}

\vfill

\paragraph*{AI use statement}
AI tools were used solely to help with the images for the lower bound. All mathematical ideas and proofs presented in this work are the authors' own. The authors take full responsibility for the content of the submitted manuscript.

\newpage
\thispagestyle{empty}
\tableofcontents

\newpage
\setcounter{page}{1}

\section{Introduction}

The problem of estimating the average degree of a graph is a fundamental problem in sublinear-time graph algorithms~\cite{Feige/SIAMJComp/2006/AverageDegree,GoldreichRon/RSA/2008/ApproximatingAvgParamGraphs,Goldreich_2017_PTBook}. Consider $\graph = \fbrac{\vertexset,\edgeset}$ to be a simple, unweighted, undirected graph with $\vertexcount$ vertices, and $\edgecount$ edges. Given a vertex $\vertex \in \vertexset$, we write its set of neighbours as $\neighbour{\vertex}$, and its degree as $\degree{\vertex} = \size{\neighbour{\vertex}}$. We denote $\degree{\edge}$ to be the degree of the edge $\edge = \fbrac{\altvertex,\vertex}$ defined as $\degree{\fbrac{\altvertex,\vertex}} = \min\fbrac{\degree{\altvertex},\degree{\vertex}}$. The average degree of the graph is $\degree{} \coloneq \nicefrac{1}{\vertexcount} \sum_{\vertex \in \vertexset}\degree{\vertex} = \nicefrac{2\edgecount}{\vertexcount}$. The arboricity of a graph $\graph = \fbrac{\vertexset,\edgeset}$, denoted by $\arboricity$, is the minimum number of spanning forests that the set of edges $\edgeset$ can be partitioned into. The goal is to design a randomized algorithm with only query access to $\graph$ to estimate $\degest$ such that $\degest \in \fbrac{1\pm\approxerror}\degree{}$, for some $\approxerror \in \fbrac{0,1}$. We say $\degest$ is a $(1\pm\approxerror)$-multiplicative approximation of $\degree{}$ if it satisfies this condition. 
For sublinear-time graph algorithms, the graph can be accessed by various queries to the graph. We begin by describing the queries that will be relevant to our work:

\begin{itemize}
    \item{\degreeq{$(\vertex)$}:} Given a vertex $\vertex$, this query returns $\degree{\vertex}$. We assume this query can be made on a $\vertex \in \vertexset$ uniformly at random. In that case, the query is called with no argument, i.e. as just \degreeq{}.
    \item{\neighbourq{$(\vertex,i)$}:} Given a vertex $\vertex \in \vertexset$ and $i \in [n]$, this query returns the $i$-th vertex $\altvertex \in \neighbour{\vertex}$ if it exists; otherwise, we get $\perp$.
    \item{\edgeexistsq{$(\altvertex,\vertex)$}:} Given two vertices $\altvertex,\vertex \in \vertexset$, this query returns $1$ if $(\altvertex,\vertex) \in \edgeset$, and $0$, otherwise.
    \item{\randedgeq{}:} This query returns an edge $\edge \in \edgeset$ uniformly at random.
    \item{\fullnbrq{$(\vertex)$}:} Given a vertex $\vertex \in \vertexset$, this query returns  the entire $\neighbour{\vertex}$
\end{itemize}

The first three queries are collectively known as \emph{local queries}, and have been extensively used in the sublinear-time graph algorithms~\cite{EdenLRS/SIAMJComp/2017/SublinearTriangle, Goldreich_2017_PTBook,  assadi2018simple, EdenRS/SODA/2020/kCliquesArboricity}. The \randedgeq{} query has recently found a lot of interest~\cite{Aliakbarpour/Algorithmica/2018/StarSubgraphEdgeSampling,assadi2018simple,BerettaTetek/TALG/2024/BetterSumEstimationViaWeightedSampling,TetekThorup/STOC/2022/EdgeSamplingFullNbrhood,bishnu2025arboricityrandomedgequeries,BerettaCS/ArXiv/2025/FasterEdgeEstimation}. The \fullnbrq{} query has been studied in the context of average degree estimation~\cite{TetekThorup/STOC/2022/EdgeSamplingFullNbrhood,BerettaCS/ArXiv/2025/FasterEdgeEstimation}. Note that \fullnbrq{} query subsumes both the \edgeexistsq{}, and \neighbourq{} queries.

Existing work on average degree estimation can be broadly categorized along two perspectives: the type of queries allowed, and the parameters governing the query complexity. The first work of estimating the average degree in sublinear-time model is due to Feige~\cite{Feige/SIAMJComp/2006/AverageDegree}, who used $\bigoten{\sqrt{\nicefrac{\vertexcount}{\degree{}}}}$ \degreeq{} queries to obtain a $(2\pm\approxerror)$-multiplicative approximation, and established that any algorithm that uses only \degreeq{} queries must make $\bigomega{\vertexcount}$ queries to obtain a $(1\pm\approxerror)$-multiplicative approximation to the average degree. Goldreich and Ron~\cite{GoldreichRon/RSA/2008/ApproximatingAvgParamGraphs} circumvented this bound using additional access to \neighbourq{} queries. They showed that using \degreeq{} and \neighbourq{} queries gives a $\bigoten{\sqrt{\nicefrac{\vertexcount}{\degree{}}}}$-query algorithm to obtain a $(1\pm\approxerror)$-multiplicative estimate of the average degree. The bound was shown to be almost (up to $\polylogneps$ factors) optimal. Motivated by the weighted sampling framework, Beretta and T\v{e}tek~\cite{BerettaTetek/TALG/2024/BetterSumEstimationViaWeightedSampling}, Motwani et al.~\cite{motwani07} established algorithms that use $\bigoten{\vertexcount^{1/3}}$ \degreeq{} and \randedgeq{} queries. The problem of estimating average degree has been studied in terms of various set membership queries \cite{harpeled18,Xichen19,BhattacharyaBGM24,addanki_et_al:LIPIcs.ESA.2022.2,RonTsur/ToCT/2016/PowerOfAnExample}. More recently, Adar and Levi~\cite{AdarL/ArXiv/2026/EdgeSampLocalIS} , Adar et al.~\cite{AdarHL/ArXiV/2026/EdgeCountLocalIS} have studied the power of the set membership queries in conjunction with the local queries.

For a more fine-grained understanding of the complexity of the problems, the graph parameter arboricity, denoted $\arboricity$, has been used to parametrize query complexity in sublinear-time settings~\cite{EdenRR/ICALP/2019/ArboricitySamplingEdges,EdenRS/SODA/2020/kCliquesArboricity, bishnu2025arboricityrandomedgequeries, Goldreich/OpenProbsGraphPT}, as well as in other models~\cite{ChibaN/SIAMJComp/1985/ArboricityandSubgraphListing,BeraSeshadriStreamingDegeneracy,LiuSeshadri/PODC/2024/MPCArboricityTriangle}. These studies are motivated by the fact that many real world graphs have low arboricity~\cite{GoelGustedtBoundedArboricity,DanischBS/WWW18.RealWorldDegeneracy,EdenRS/SODA/2020/kCliquesArboricity}. In the context of estimating average degree, Eden et al.~\cite{EdenRS/SIAMJDM/2019/SublinearDegDistMoments} obtained an algorithm that uses $\bigoten{\nicefrac{\arboricity}{\degree{}}}$ \degreeq{} and \neighbourq{} queries, and established tight lower bounds in terms of $\arboricity$ and $\degree{}$. Recently, Eden and Seshadri~\cite{EdenS/ArXiv/2026/NoteAvgDegArboricity} provided a tighter analysis to improve the bound to $\bigo{\nicefrac{\arboricity}{\approxerror^2\degree{}}}$ queries. This parametrization on $\arboricity$ is interesting as $\nicefrac{\arboricity}{\degree{}}$ is always smaller than $\sqrt{\nicefrac{\vertexcount}{\degree{}}}$, and it can be much smaller for low-arboricity graphs. However, this algorithm needs the knowledge of $\vertexcount$, as well as a constant factor upper bound on arboricity $\arboricity$. They established that any algorithm that makes only \degreeq{} and \neighbourq{} queries must make $\bigomega{\sqrt{\vertexcount}}$ queries without an advice on the arboricity $\arboricity$. Eden et al.~\cite{EdenRV/ArXiV/2025/TestableAlgoTrianglEdges} showed that if \randedgeq{} queries are allowed in addition, the same upper bound can be attained without the advice on $\arboricity$.

All the algorithms discussed above assume the knowledge of the number of vertices $\vertexcount$. A complementary line of work considers parameter-oblivious algorithms, where even the graph parameters number of vertices $\vertexcount$, and arboricity $\arboricity$ is unknown. Goldreich~\cite{Goldreich/OpenProbsGraphPT} noted that designing such algorithms is an important question in sublinear computation. In this setting, Beretta and T\v{e}tek~\cite{BerettaTetek/TALG/2024/BetterSumEstimationViaWeightedSampling} first established an algorithm that estimates the average degree using $\bigo{\nicefrac{\sqrt{\vertexcount}}{\approxerror}+\nicefrac{\log\vertexcount}{\approxerror^2}}$ \degreeq{} and \randedgeq{} queries. They established a matching lower bound for the general sum-estimation problem where $\vertexcount$ is unknown. Recently, Beretta et al.~\cite{BerettaCS/ArXiv/2025/FasterEdgeEstimation} established the first graph-specific algorithm to estimate average degree when $\vertexcount$ is unknown. They showed that even without any knowledge of $\vertexcount$, when \neighbourq{} is allowed in addition to the \degreeq{}, and \randedgeq{} queries, average degree can be estimated using $\bigoten{\sqrt{\nicefrac{\vertexcount}{\degree{}}}}$ queries. They showed that when $\vertexcount$ is known, allowing  \edgeexistsq{} queries reduces the number of queries to $\bigoten{\sqrt[3]{\nicefrac{\vertexcount}{\degree{}}}}$, and allowing \fullnbrq{} further reduces it to $\bigoten{\sqrt[4]{\nicefrac{\vertexcount}{\degree{}}}}$. They established lower bounds showing that these bounds are tight in terms of $\vertexcount$ and $\degree{}$. However, their algorithms do not achieve arboricity-dependent query complexity, and obtaining such bounds was posed as an open question in their work. Motivated by their questions, we study the following:

\begin{blackbox}
\begin{center}
    \textit{Can average degree be estimated parameter-obliviously
    with $\arboricity$-sensitive complexity?\\
    and\\
    What power do different queries provide for this problem?}
\end{center}
\end{blackbox}

We answer the first question in the affirmative. In fact, perhaps surprisingly, we show that an algorithm that makes only $\bigot{\nicefrac{1}{\approxerror^2}}$ \randedgeq{} queries in addition to $\bigot{\nicefrac{\arboricity}{\approxerror^2\degree{}}}$ \degreeq{} queries achieves this. Regarding the second question, for the case of unknown $\vertexcount$, we show that this bound is tight with respect to graph parameters, even with access to \neighbourq{}, \edgeexistsq{}, and \fullnbrq{} queries. Furthermore, even when $\vertexcount$ is known, our lower bounds show even with all the queries discussed above, no better bounds can be achieved for all ranges of $\arboricity$. We also establish $\approxerror$-sensitive lower bounds showing that $\bigomega{\nicefrac{1}{\approxerror}}$ \randedgeq{} queries are necessary to obtain a sublinear time algorithm using \degreeq{} and \randedgeq{} queries. Furthermore, complementing the results of Eden and Seshadri~\cite{EdenS/ArXiv/2026/NoteAvgDegArboricity}, we settle the arboricity-sensitive query complexity for an algorithm using \degreeq{}, and \fullnbrq{} queries.

The remainder of the paper is organized as follows. \Cref{Sec: Results} presents our main results and discusses their implications. \Cref{Section: TechOverview} provides an overview of the key ideas and techniques underlying our work. \Cref{Section: Prelims} introduces the necessary preliminaries. \Cref{Sec: Algorithm} describes our main algorithmic contribution. Finally, \Cref{Sec: Lower Bounds} establishes the corresponding lower bounds.

\section{Overview of Our Results}\label{Sec: Results}


In this section, we state the main results in this work. Throughout this paper, we assume that $\edgecount = \bigomega{\vertexcount}$. Such assumptions are standard in the literature~\cite{GoldreichRon/RSA/2008/ApproximatingAvgParamGraphs,EdenRS/SIAMJDM/2019/SublinearDegDistMoments,BerettaTetek/TALG/2024/BetterSumEstimationViaWeightedSampling,BerettaCS/ArXiv/2025/FasterEdgeEstimation}, and made to facilitate a simpler exposition. When this assumption is violated, given access to \randedgeq{} queries, Ron and Tsur~\cite{RonTsur/ToCT/2016/PowerOfAnExample} provide a collision-based estimator that uses $\smallo{\sqrt{\vertexcount}}$ queries. All our upper bounds hold in the size-oblivious setting, i.e., the algorithms do not require prior knowledge of the number of vertices $n$. We begin with our main algorithmic contribution.

\begin{blackbox}
\begin{restatable}[Upper Bound - General Graphs]{theorem}{UBG}\label{Thm: Main UB - General}
    There exists an algorithm that, given access to \degreeq{} and \randedgeq{} queries to a graph $\graph$ with average degree $\degree{}$, outputs an  estimate $\degest$ such that $\degest \in (1\pm\approxerror)\degree{}$ with probability at least $\frac{9}{10}$,  and makes $\bigot{\frac{\arboricity}{\approxerror^2\degree{}}}$ \degreeq{}, and $\bigot{\frac{1}{\approxerror^2}}$ \randedgeq{} queries in expectation, where $\arboricity$ is the arboricity of the graph.
\end{restatable}
\end{blackbox}

\begin{table}[ht]
    \renewcommand{\arraystretch}{1.6}
    \centering
    \begin{tabular}{c|c|c|c|c|c|c|c}
    \hline
     & \cite{GoldreichRon/RSA/2008/ApproximatingAvgParamGraphs} & \cite{EdenRS/SIAMJDM/2019/SublinearDegDistMoments,EdenS/ArXiv/2026/NoteAvgDegArboricity} &  \cite{BerettaTetek/TALG/2024/BetterSumEstimationViaWeightedSampling} & \cite{EdenRV/ArXiV/2025/TestableAlgoTrianglEdges} & \multicolumn{2}{|c|}{\cite{BerettaCS/ArXiv/2025/FasterEdgeEstimation}}  & {\Cref{Thm: Main UB - General}}\\ \hline
    \degreeq{} & $\bigo{\sqrt{\frac{\vertexcount}{\degree{}}}}$& $\bigo{{\frac{\arboricity}{\degree{}}}}$ & $\bigo{\degree{}}$ & $\bigo{{\frac{\arboricity}{\degree{}}}}$ & $\bigo{\sqrt{\frac{\vertexcount}{\degree{}}}}$ & $\bigo{\sqrt[3]{\frac{\vertexcount}{\degree{}}}}$ & {$\bigo{\frac{\arboricity}{\degree{}}}$} \\ \hline
    \neighbourq{} & $\bigo{\sqrt{\frac{\vertexcount}{\degree{}}}}$ & $\bigo{{\frac{\arboricity}{\degree{}}}}$ & \emptycell & $\bigo{{\frac{\arboricity}{\degree{}}}}$ & $\bigo{\sqrt{\frac{\vertexcount}{\degree{}}}}$ & $\bigo{1}$ & \emptycell \\ \hline
    \randedgeq{} & \emptycell & \emptycell & $\bigo{\degree{}}$ & $\bigo{1}$ & $\bigo{1}$ & $\bigo{\sqrt[3]{\frac{\vertexcount}{\degree{}}}}$ & $\bigo{1}$ \\ \hline
    \edgeexistsq{} & \emptycell & \emptycell & \emptycell & \emptycell & \emptycell & $\bigo{\sqrt[3]{\frac{\vertexcount}{\degree{}}}}$ & \emptycell \\ \hline

    Known Input & \multirow{2}{*}{$\vertexcount$} & \multirow{2}{*}{$\arboricity$} & \multirow{2}{*}{None} & \multirow{2}{*}{$\vertexcount$} & \multirow{2}{*}{None} & \multirow{2}{*}{$\vertexcount$} & \multirow{2}{*}{None} \\
    Parameters & & & & & & & \\ \hline
    \end{tabular}
    \caption{Overview of query complexities for estimating the average degree across previous literature and \Cref{Thm: Main UB - General}. We suppress the $\poly{\approxerror^{-1},\log\vertexcount}$ terms in this table for presentation purposes.}
    \label{tab:AsymptoticAdvantage}
\end{table}
\begin{remark}
    \textbf{Asymptotically smaller number of \randedgeq{} queries:} Since the work of Goldreich and Ron~\cite{GoldreichRon/RSA/2008/ApproximatingAvgParamGraphs}, all algorithms that achieve a $(1\pm\approxerror)$-multiplicative estimate using sublinear ($\smallo{\nicefrac{\vertexcount}{\degree{}}}$) \degreeq{} queries also rely on asymptotically same number of either \neighbourq{} or \randedgeq{} queries (See \Cref{tab:AsymptoticAdvantage}). 
    \Cref{Thm: Main UB - General} shows that only $\bigo{\nicefrac{1}{\approxerror^2}}$ \randedgeq{} queries suffice to achieve such guarantees when combined with \degreeq{} queries. Leveraging this fact with existing results on edge sampling~\cite{eden2022samplingedgesublineartime,EdenRR/ICALP/2019/ArboricitySamplingEdges} matches the existing algorithms using \degreeq{} and \neighbourq{} queries up to $\poly{\approxerror^{-1}\log\vertexcount}$ factor when graph parameters are known. Details are presented in \Cref{App: SimEdge}.
\end{remark}

Note that unlike many other algorithms for edge estimation~\cite{GoldreichRon/RSA/2008/ApproximatingAvgParamGraphs,EdenRS/SIAMJDM/2019/SublinearDegDistMoments,BerettaCS/ArXiv/2025/FasterEdgeEstimation}, we do not need to assume any lower bound on the degree of individual vertices for \Cref{Thm: Main UB - General} to hold. However, given a guarantee that a graph does not have isolated vertices, we can improve this bound using a subroutine of the algorithms presented in earlier works~\cite{BerettaTetek/TALG/2024/BetterSumEstimationViaWeightedSampling,BerettaCS/ArXiv/2025/FasterEdgeEstimation}.

\begin{blackbox}
    \begin{restatable}[Upper Bound - No Isolated Vertices]{theorem}{UBNI}\label{Thm: Main UB - No Isolated Vertices}
    There exists an algorithm that, given access to \degreeq{} and \randedgeq{} queries to a graph $\graph$ with average degree $\degree{}$, and no isolated vertices outputs an  estimate $\degest$ such that $\degest \in (1\pm\approxerror)\degree{}$ with probability at least $\frac{9}{10}$, and makes $\bigot{\min\fbrac{\frac{\degree{}}{\approxerror^2},\frac{\arboricity}{\approxerror^2\degree{}}}}$ queries in expectation, where $\arboricity$ is the arboricity of the graph.
\end{restatable}
\end{blackbox}

\begin{remark}
    \textbf{Regarding upper bounds:} Our algorithm, despite having no prior knowledge of any graph parameters, consistently matches or outperforms the best-known algorithms across a range of query accesses with the knowledge of $\vertexcount$. In particular, it performs at least as well as algorithms that has access to \degreeq{}, \neighbourq{}, and \edgeexistsq{} queries \cite{Feige/SIAMJComp/2006/AverageDegree,GoldreichRon/RSA/2008/ApproximatingAvgParamGraphs,EdenRosenbaum/Approx/2018/LowerBoundGraphCommunication}, those employing \degreeq{} and \randedgeq{} queries \cite{BerettaTetek/TALG/2024/BetterSumEstimationViaWeightedSampling,motwani07}, and those relying on Independent Set queries~\cite{ChenLW/SODA/2020/OptimalISOracles}. 
\end{remark}

\begin{table}[ht]
\centering
\renewcommand{\arraystretch}{1.6}
\begin{tabular}{|c|c|c|c|c|}
\hline
\textbf{$\vertexcount$} & Queries Allowed & \textbf{$\arboricity \leq \vertexcount^{2/5}$} & \textbf{$\arboricity \in [ \vertexcount^{2/5},\sqrt{\vertexcount}]$} & \textbf{$\sqrt{\vertexcount} \leq \arboricity$} \\ \hline


Not & $\degreeq{}$ &
\multicolumn{3}{|c|}{\cellcolor{Cyan!20}$\bigot{\min\fbrac{\frac{\degree{}}{\approxerror^2},\frac{\arboricity}{\approxerror^2\degree{}}}}$} \\

Known &  + \randedgeq{} &
\multicolumn{3}{|c|}{\cellcolor{Cyan!20}[\Cref{Thm: Main UB - No Isolated Vertices}]} \\ \hline

Known & + \neighbourq{}  &  \multicolumn{3}{|c|}{\cellcolor{Cyan!20}$\bigomega{\min\fbrac{\degree{},\frac{\arboricity}{\degree{}}}}$~~~[\Cref{Thm: Lower Bound}]} \\ \hline

\multirow{2}{*}{Known} & \multirow{2}{*}{+ \edgeexistsq{} } &  
\multicolumn{2}{|c|}{\cellcolor{Cyan!20}$\bigomega{\min\fbrac{\degree{},\frac{\arboricity}{\degree{}}}}$}  & $\bigoten{\min\fbrac{\degree{},\sqrt[3]{\frac{\vertexcount}{\degree{}}}}}$~\cite{BerettaCS/ArXiv/2025/FasterEdgeEstimation}\\ 
&  & \multicolumn{2}{|c|}{\cellcolor{Cyan!20}[\Cref{Thm: Lower Bound}]} & $\bigomega{\min\fbrac{\degree{},\sqrt[3]{\frac{\vertexcount}{\degree{}}}}}~\cite{BerettaCS/ArXiv/2025/FasterEdgeEstimation}$ \\ \hline

\multirow{2}{*}{Known} & \multirow{2}{*}{+ \fullnbrq{}} & \multicolumn{1}{|c|} {\cellcolor{Cyan!20}$\bigomega{\min\fbrac{\degree{},\frac{\arboricity}{\degree{}}}}$} & 
\multicolumn{2}{c|}{$\bigoten{\min\fbrac{\degree{},\sqrt[4]{\frac{\vertexcount}{\degree{}}}}}$~\cite{BerettaCS/ArXiv/2025/FasterEdgeEstimation}}\\
&  & \multicolumn{1}{|c|}{\cellcolor{Cyan!20}[\Cref{Thm: Lower Bound}]} & 
\multicolumn{2}{c|}{$\bigomega{\min\fbrac{\degree{},\sqrt[4]{\frac{\vertexcount}{\degree{}}}}}$~\cite{BerettaCS/ArXiv/2025/FasterEdgeEstimation}}\\ \hline


Not & All of the & \multicolumn{3}{|c|}{\cellcolor{Cyan!20}$\bigomega{\min\fbrac{\degree{},\frac{\arboricity}{\degree{}}}}$}\\
Known & Above & \multicolumn{3}{|c|}{\cellcolor{Cyan!20}[\Cref{Thm: Lower Bound Unknown N}]}\\\hline

\end{tabular}

\caption{Summary of the query complexities of estimating the average degree across various ranges of arboricity for the queries considered in this work. The cells highlighted in \textcolor{Cyan!60}{Blue} contain results presented in this paper. The other bounds are due to~Beretta et al.~\cite{BerettaCS/ArXiv/2025/FasterEdgeEstimation}. Note that '+' indicates access to the query in addition to all queries specified in the previous rows.}
\label{Table: Results}
\end{table}

Next, we state the lower bound results of this work. The first result establishes lower bounds even when the graph parameters $\vertexcount$ and $\arboricity$ is known to the algorithm. It considers a hierarchy of query accesses, and establishes ranges of arboricity $\arboricity$ where our algorithm is optimal. We say a set of $\vertexcount$,  $\arboricity$, and $\degree{}$ are \emph{valid} if there exists a graph with $\vertexcount$ vertices, average degree $\degree{}$, and arboricity $\arboricity$.

\begin{blackbox}
    \begin{restatable}[Lower Bound: $\vertexcount$ known]{theorem}{LB}\label{Thm: Lower Bound}
    Consider valid $\vertexcount$,  $\arboricity$, and $\degree{}$ such that $\nicefrac{\arboricity}{4} \geq \degree{} \geq 4$. Then, there exists a graph with $\vertexcount$ vertices, average degree $\degree{}$, arboricity $\arboricity$, and no isolated vertices such that any algorithm that can obtain a $(1\pm\approxerror)$-multiplicative approximation of the average degree $\degree{}$ of the graph with probability at least $\frac{3}{4}$ for any $\approxerror < \frac{1}{3}$,
    \begin{enumerate}
        \item using \degreeq{}, \neighbourq{}, and \randedgeq{} queries, must make at least\\ $\bigomega{\min\fbrac{\degree{},\frac{\arboricity}{\degree{}}}}$ queries.
        \item using \degreeq{}, \neighbourq{}, \edgeexistsq{} and \randedgeq{} queries, must make at least \\$\bigomega{\min\fbrac{\degree{},\frac{\arboricity}{\degree{}}}}$ queries, given $\arboricity \leq \sqrt{\vertexcount}$.
        \item using \degreeq{}, \neighbourq{}, \edgeexistsq{}, \fullnbrq{} and \randedgeq{} queries, must make at least $\bigomega{\min\fbrac{\degree{},\frac{\arboricity}{\degree{}}}}$ queries, given $\arboricity \leq \vertexcount^{2/5}$.
    \end{enumerate}
\end{restatable}
\end{blackbox}

We next consider the setting where the number of vertices $n$ is unknown. In this scenario, we show that our upper bounds are optimal for all ranges of arboricity $\arboricity$, even against algorithms with access to the full spectrum of queries. 

\begin{blackbox}
    \begin{restatable}[Lower Bound: $\vertexcount$ not known]{theorem}{LBUnknownN}\label{Thm: Lower Bound Unknown N}
    Consider valid $\vertexcount$,  $\arboricity$, and $\degree{}$ such that $\nicefrac{\arboricity}{4} \geq \degree{} \geq 4$. Then there exists a graph with $\vertexcount$ vertices, average degree $\degree{}$, and arboricity $\arboricity$ such that any algorithm that can obtain a $(1\pm\approxerror)$-multiplicative approximation of the average degree $\degree{}$ with probability at least $\frac{3}{4}$ for any $\approxerror \leq \frac{1}{4}$ without the knowledge of $\vertexcount$, and using \degreeq{}, \neighbourq{}, \edgeexistsq{}, \fullnbrq{}, and \randedgeq{} must make $\bigomega{\min\fbrac{\degree{},\frac{\arboricity}{\degree{}}}}$ queries.
\end{restatable}
\end{blackbox}

\begin{remark}
    \textbf{Regarding graph-parameter-based lower bounds:} Our lower bounds characterize the complexity of edge estimation across different query models in terms of the arboricity $\arboricity$. When $\vertexcount$ is known, we provide tight bounds for algorithms using \degreeq{}, \neighbourq{}, and \randedgeq{} queries for all values of $\arboricity$; for those additionally using \edgeexistsq{} queries when $\arboricity \leq \sqrt{\vertexcount}$; and for those further using \fullnbrq{} queries when $\arboricity \leq \vertexcount^{2/5}$. When $\vertexcount$ is unknown, we establish tight bounds for algorithms with access to all the queries mentioned above.
\end{remark}

Next, we establish $\approxerror, \confidence$-sensitive lower bounds.

\begin{blackbox}
    \begin{restatable}[$\approxerror,\confidence$-Sensitive Lower bound]{theorem}{LBEpsDel}\label{Theorem: Lower Bound Epsilon Delta}
    Consider valid $\vertexcount$,  $\arboricity$, and $\degree{}$ such that $\arboricity \geq 12 \degree{}$. Then, there exists a graph with $\vertexcount$ vertices, average degree $\degree{}$, and arboricity $\arboricity$ such that any algorithm that makes $\smallo{\frac{1}{\approxerror^2}}$ \randedgeq{} queries and $\smallo{\frac{\arboricity}{\approxerror^2\degree{}}\log\fbrac{\frac{1}{\confidence}}}$ \degreeq{}, \neighbourq{}, \edgeexistsq{}, and \fullnbrq{} queries can not obtain a $(1\pm\approxerror)$-multiplicative approximation of the average degree $\degree{}$ of the graph with probability at least $1 - \confidence$ for any $\frac{1}{8} \geq \approxerror \geq \frac{4\arboricity}{\sqrt{\edgecount}}\sqrt{\log\frac{1}{\confidence}}$.
\end{restatable}
\end{blackbox}

\begin{remark}
    \textbf{Error dependence in $\arboricity$-parameterized bounds}: We show that parameterizing the query complexity in terms of $\arboricity$ incurs an additional dependence on $\nicefrac{1}{\approxerror}$. 
    In contrast, T\v{e}tek and Thorup~\cite{TetekThorup/STOC/2022/EdgeSamplingFullNbrhood} show that with access to \degreeq{} and \fullnbrq{} queries, the complexity of estimating the average degree is 
    $
    \widetilde{\Theta}\!\left(\frac{1}{\approxerror}\sqrt{\frac{\vertexcount}{\degree{}}}\right).
    $
    
    Note that the lower bound on $\approxerror$ in \Cref{Theorem: Lower Bound Epsilon Delta} is required as otherwise the result mentioned above provides a better algorithm. Combining \Cref{Theorem: Lower Bound Epsilon Delta} with the upper bound of Eden and Seshadri~\cite{EdenS/ArXiv/2026/NoteAvgDegArboricity}, we obtain that the optimal arboricity-dependent query complexity with the same query access (\degreeq{} and \fullnbrq{}) is 
    $
    \Theta\!\left(\frac{\arboricity}{\approxerror^2 \degree{}}\right),
    $
    thereby requiring an additional $\nicefrac{1}{\approxerror}$ factor.  This additional dependency on $\approxerror$ have also been observed in arboricity-sensitive algorithms for edge sampling \cite{eden2022samplingedgesublineartime,EdenRR/ICALP/2019/ArboricitySamplingEdges} and triangle counting \cite{bishnu2025arboricityrandomedgequeries, DBLP:conf/approx/AssadiN22}. Understanding whether this additional dependence on $\approxerror$ is inherent to arboricity parametrization for various problems is an important research question.
\end{remark}

The next statement establishes the need of \randedgeq{} queries for the problem of estimating average degree. In particular, it shows that any algorithm that circumvents the lower bound of Feige~\cite{Feige/SIAMJComp/2006/AverageDegree} must make $\bigomega{\nicefrac{1}{\approxerror}}$ \randedgeq{} queries.

\begin{blackbox}
    \begin{restatable}[Lower Bound - \randedgeq{} and \degreeq{}]{theorem}{LBFeige}\label{Thm: LBFeige}
    Consider valid $\vertexcount,\arboricity$ , and $\degree{}$ such that $\degree{}\leq \frac{\arboricity}{4}$, and $\approxerror \leq \frac{1}{4}$. Then there exists a graph with $\vertexcount$ vertices, average degree $\degree{}$, and arboricity $\arboricity$ such that any algorithm that can obtain a $(1\pm\approxerror)$-multiplicative approximation of the average degree $\degree{}$ with probability at least $\frac{3}{4}$, and makes $\smallo{\frac{1}{\approxerror}}$ \randedgeq{} queries must make $\bigomega{\frac{\vertexcount}{\approxerror\degree{}}}$ \degreeq{} queries.
\end{restatable}
\end{blackbox}

\begin{table}[ht]
    \renewcommand{\arraystretch}{1.4}
    \centering
    \begin{tabular}{cccc}
    \hline

    Number of & Number of & Other Queries &  \\
    \randedgeq{} Queries & Other Queries & Allowed  & Reference\\\hline

    $\smallo{\frac{1}{\approxerror}}$ 
    & $\bigomega{\frac{\vertexcount}{\approxerror\degree{}}}$ & \degreeq{}
    & [\Cref{Thm: LBFeige}] \\\hline
    
    $\bigot{\frac{1}{\approxerror^2}}$ 
    & $\bigot{\frac{\arboricity}{\approxerror^2\degree{}}}$ & \degreeq{} 
    & [\Cref{Thm: Main UB - General}] \\\hline

    $\smallo{\frac{1}{\approxerror^2}}$ 
    & $\bigomega{\frac{\arboricity}{\approxerror^2\degree{}}}$ & \degreeq{} + \fullnbrq{}
    & [\Cref{Theorem: Lower Bound Epsilon Delta}] \\\hline

    \end{tabular}
    \caption{Summary of $\approxerror$-sensitive bounds for different amounts of \randedgeq{} queries.}
    \label{tab:RandEdgeSplit}
\end{table}

\begin{remark}
    \textbf{Power of \randedgeq{} Queries:} Our results (\Cref{tab:RandEdgeSplit}) characterize the role of \randedgeq{} queries in enabling sublinear-time estimation of the average degree. In particular, \Cref{Thm: Main UB - General} establishes that only $\bigo{\nicefrac{1}{\approxerror^2}}$ \randedgeq{} queries are sufficient to achieve sublinear number of $\degreeq{}$ queries. Complementing this, \Cref{Thm: LBFeige} shows that any algorithm using $\smallo{\nicefrac{\vertexcount}{\degree{}}}$ \degreeq{} queries must make at least $\Omega\!\left(\nicefrac{1}{\approxerror}\right)$ \randedgeq{} queries.
    
\end{remark}

Some further implications of our results are discussed in \Cref{SubSec: Discussion}.

\section{Overview of Techniques and Key Ideas}\label{Section: TechOverview}

In this section, we give a brief overview of the technical challenges and ideas of this work. We set up notations in~\Cref{SubSec: Notations}, describe the ideas behind the upper bounds in~\cref{SubSec: UB Overview}, describe the ideas behind the lower bounds in~\Cref{SubSec: LB Overview}.

\subsection{Setup and Notation}\label{SubSec: Notations}

We introduce notations that capture the key structural decomposition used by our algorithm: separating vertices (and edges) into low-degree and high-degree parts. This decomposition enables variance control for mean estimation and plays a central role in both the estimator and its analysis. Throughout the paper, we denote $\constant$ to be a large enough universal constant.  A central idea in our approach is to truncate high-degree vertices to control variance. Given a threshold $\threshold$, we define vertices to be \emph{heavy} or \emph{light} depending on whether their degrees are higher or lower than the given threshold $\threshold$, respectively.

\begin{definition}[Heavy and Light Vertices - $(\threshold)$]\label{Def: Heavy Light Vertices}
    Given a threshold $\threshold$, we say a vertex $\vertex$ is heavy if $\degree{\vertex} > \threshold$, and light if $\degree{\vertex} \leq \threshold$.
\end{definition}

In this work, the threshold $\threshold$ is clear from the context. We denote the set of heavy (resp. light) vertices as $\heavyverts$ (resp. $\lightverts$). Formally, we have:
\begin{align*}
    &\heavyverts = \sbrac{\vertex \in \vertexset|\degree{\vertex} > \threshold} &\text{and}&&\lightverts = \sbrac{\vertex \in \vertexset|\degree{\vertex} \leq \threshold}
\end{align*}

Note that the sets $\heavyverts$, and $\lightverts$ is a partition of the set of vertices $\vertexset$. We denote $\hedgecnt$, and $\ledgecnt$ to be the sums of degrees of the heavy and light vertices, respectively. Formally, we have:
\begin{align*}
    &\hedgecnt = \sum_{\heavyvert \in \heavyverts} \degree{\heavyvert} &\text{and}&&\ledgecnt = \sum_{\lightvert \in \lightverts} \degree{\lightvert}
\end{align*}

 We also denote $\heavyedges,\lightedges$, and $\hledges$ to be the set of edges between $\heavyvert_i, \heavyvert_j \in \heavyverts$,  $\lightvert_i,\lightvert_j \in \lightverts$, and $\heavyvert_i \in \heavyverts,\lightvert_j \in \lightverts$, respectively. Based on these notations, we have:
    \begin{align*}
         &\hedgecnt = 2\size{\heavyedges} + \size{\hledges} &\text{and}&&\ledgecnt = 2\size{\lightedges} + \size{\hledges}
    \end{align*}

We denote $\dltrue$ to be the contribution of light vertices to the
average degree, i.e., $$\dltrue = \frac{\ledgecnt}{\vertexcount}.$$
We denote $\ldens$ to be the fraction of edges incident to light vertices, i.e., $$\ldens = \frac{\ledgecnt}{2\edgecount}.$$
Thus, one can think of $\ldens$ as the density of the light edges. We now define a \emph{good} threshold property, that we will leverage in our algorithm.

\begin{definition}[Good Threshold]\label{Def: Good Threshold}
    We define a threshold $\threshold$ to be good if $\ledgecnt \geq \frac{\edgecount}{2}$, and consequently $\ldens \geq \frac{1}{4}$.
\end{definition}

\subsection{Upper Bound}\label{SubSec: UB Overview}
Estimating the average degree can be viewed as a mean estimation problem. However, a direct application of standard techniques fails due to the presence of high-degree vertices, which induce large variance. A standard approach to such issues in mean estimation is to bound the variance by setting a maximum value that any sample can take. In our setting, this corresponds to only considering vertices with degree less than a threshold $\threshold$. 

However, this results in an underestimation due to the high degree vertices being ignored. Given a good threshold $\threshold$, we resolve this issue using a ratio of estimators technique by combining a truncated mean estimator with a coin bias estimator. We also leverage a modified version of the coin bias estimator to find a good threshold for our purposes. An overview of this approach is illustrated in \Cref{Figure: Blockdiag}. We outline the broad overview of our algorithm below:

\noindent\textbf{Ratio of Estimators:} We start with the observation that the average degree can be decomposed as:
\[
\degree{} = \frac{2 \edgecount}{\vertexcount} = \frac{\ledgecnt / \vertexcount}{\ledgecnt / 2 \edgecount} = \frac{\dltrue}{\ldens}
\]
The average degree can be seen as the ratio of $\dltrue$ and $\ldens$, both of which depend only on light vertices. If we can get the estimates of these two mean quantities $\dltrue$ and $\ldens$ and combine them to get an estimate of $\degree{}$, we can avoid the usage of any quantity that depends on the high degree vertices. Thus, it suffices to estimate the two quantities: $\dltrue$ via vertex sampling, and $\ldens$ via edge sampling, and combine them to obtain an estimate of $d$.

We start with the estimation of $\dltrue$. We define a truncated estimator by assigning value $0$ to heavy vertices and retaining the true degree for light vertices. This can be implemented using \degreeq{} queries on a random set of vertices with simple post-processing. The resulting sample mean is an unbiased estimator of $\dltrue$; denote this estimate as $\dlest$.
Due to the upper bound of $\threshold$ on the degree values of the samples considered, we can obtain a $(1\pm\approxerror)$-estimate of $\dltrue$ using $\bigo{\nicefrac{\threshold}{(\approxerror^2\dltrue)}}$ $\degreeq{}$ queries by Chebyshev's inequality.

Next, we describe the estimation of $\ldens$. Observe that the definition of the good threshold ensures that $\ldens = \nicefrac{\ledgecnt}{2\edgecount} \geq \nicefrac{1}{4}$. Hence, if we can simulate a coin toss with bias $\ldens$, we can obtain a $(1\pm\approxerror)$-multiplicative approximation using $\bigot{\nicefrac{1}{\approxerror^2}}$ tosses. We simulate this coin toss using \randedgeq{} queries in conjunction with \degreeq{} queries: sample a random edge, choose a uniformly random endpoint, and output $1$ if it is light and $0$ otherwise. Since each edge is sampled uniformly, and consequently each endpoint is chosen uniformly,
the probability of observing a light endpoint is exactly proportional to the
sum of degrees of light vertices, i.e., $\ldens = \nicefrac{\ledgecnt}{2\edgecount}$. The probability that this end point is light is exactly $\ldens$. Hence, this procedure yields an unbiased estimator of $\ldens$.

\begin{figure}[ht]
\centering
\begin{tikzpicture}[
  block/.style = {rectangle, draw, thick, rounded corners, align=center,
                  minimum height=1.4cm, minimum width=3.5cm, fill=blue!5},
  io/.style    = {rectangle, draw, thick, rounded corners, align=center,
                  minimum height=1.4cm, minimum width=3.0cm, fill=orange!10},
  process/.style = {rectangle, draw, thick, rounded corners, align=center,
                    minimum height=1.4cm, minimum width=3.5cm, fill=green!8},
  smalllabel/.style = {font=\small, inner sep=1pt},
  scale = 0.85
]

\node[io] (graph) {Graph};

\node[block, right=3.5cm of graph, yshift=1.8cm] (meanest)
  {\meanest{} $(\dlest)$};

\node[block, below=1.8cm of meanest] (coinbias)
  {\ldensest{}\\ Estimation$(\estldens)$};

\node[coordinate, right=8.5cm of graph] (merge) {};

\node[process, right=0.5cm of merge] (estimate)
  {Ratio of Estimators\\ $\degest = \frac{\dlest}{\estldens}$};

\draw[->, thick] (graph.east) to[out=20,in=180]
  node[midway, above, smalllabel] {$\degreeq{}$} (meanest.west);
\draw[->, thick] (graph.east) to[out=-20,in=180]
  node[midway, above, smalllabel] {$\degreeq{}$,} node[midway, below, smalllabel] {$\randedgeq{}$} (coinbias.west);

\draw[-, thick] (meanest.east) to[out=0,in=120] (merge);
\draw[-, thick] (coinbias.east) to[out=0,in=240] (merge);

\draw[->, thick] (merge) -- (estimate.west);

\node[above=3mm of graph] {\textbf{Query Access}};

\end{tikzpicture}
\caption{Overview of the main ideas behind the algorithm.}\label{Figure: Blockdiag}
\end{figure}


\noindent\textbf{Finding a Good Threshold:} The remaining challenge is to identify a good threshold $\threshold$. For a threshold to be good, the subgraph induced by the set of heavy vertices $\heavyverts$ must contain only a small fraction of the total edges. Equivalently, a constant fraction of edges should be incident to light vertices. By a corollary of the Nash-Williams Theorem~(\Cref{Lemma: Arboricity Edge Bound}), any choice of $\threshold \ge c\arboricity$ ensures this property, implying that $\ldens = \Omega(1)$. However, this guarantee is existential: it yields an algorithm that succeeds given access to such a threshold, but does not specify how to find one.

For our purposes, any threshold $\threshold$ that satisfies the good threshold property~(\Cref{Def: Good Threshold}) is sufficient. Since $\arboricity$ is unknown, we search for such a threshold by testing candidate values of $\threshold$. For each candidate, we estimate $\ldens$ up to constant additive error and accept $\threshold$ if the estimate exceeds a fixed threshold. In particular, we accept $\threshold$ if the estimate $\ldensest$ exceeds a constant (e.g., $5/16$), which certifies that $\tau$ is a good threshold.

\begin{figure}[ht]
\centering
\begin{tikzpicture}[
  block/.style = {rectangle, draw, thick, rounded corners, align=center,
                  minimum height=1.4cm, minimum width=3cm, fill=gray!5},
  io/.style    = {rectangle, draw, thick, rounded corners, align=center,
                  minimum height=1.4cm, minimum width=3.0cm, fill=orange!10},
  process/.style = {rectangle, draw, thick, rounded corners, align=center,
                    minimum height=1.4cm, minimum width=3cm, fill=green!8},
  smalllabel/.style = {font=\small, inner sep=1pt}
]


\node[block] (noadv)
  {\noadv};

\node[block, right=2.75cm of noadv] (thresadv)
  {\thradv};

\node[block, right=2.75cm of thresadv] (alladv)
  {\alladv};


\draw[->, thick] (noadv.east) --
  node[midway, above = 0.1cm, smalllabel] {Provides a}
  node[midway, below = 0.1cm, smalllabel] {good threshold $\threshold$}
  (thresadv.west);

\draw[->, thick] (thresadv.east) --
  node[midway, above = 0.1cm, smalllabel] {Provides a lower}
  node[midway, below = 0.1cm, smalllabel] {bound on $\degree{}$}
  (alladv.west);


\node[above=3mm of thresadv] {\small uses $\tau$};
\node[above=3mm of alladv] {\small uses $\tau, \degbnd$};


\end{tikzpicture}
\caption{The hierarchy of algorithms and their corresponding advices}\label{Figure: SearchDiagram}
\end{figure}
\noindent\textbf{Wrapping Up:} Finally, the algorithm proceeds by searching for a good threshold $\tau$, and then invoking the estimator with this choice. We use well known searching techniques to find an appropriate lower bound of average degree \cite{EdenRS/SODA/2020/kCliquesArboricity,EdenLRS/SIAMJComp/2017/SublinearTriangle}. Since larger thresholds are accepted with decreasing probability, taking expectation over the search process yields the desired $\widetilde{O}(\alpha/d)$ bound. We outline the flow of algorithms in \Cref{Figure: SearchDiagram}.

\subsection{Lower Bound}\label{SubSec: LB Overview}

Our lower bounds are based on constructing pairs of graph instances that differ in their average degree but are indistinguishable under limited query access. We outline the main ideas below.

\noindent\textbf{Lower Bounds: Graph Parameters (\Cref{SubSec: GeneralLB}).}
We construct graphs consisting of $\lbclqcnt$ disjoint cliques of size $\arboricity$ together with a matching on the remaining vertices. By varying the number of cliques (e.g., $\lbclqcnt$ vs.\ $2\lbclqcnt$), we obtain instances with different average degrees while preserving arboricity.

When the number of queries is small, all queries return locally consistent information. This yields two regimes: when $\degree{} \ge \sqrt{\arboricity}$, distinguishing requires $\Omega(\nicefrac{\arboricity}{\degree{}})$ queries, while for $\degree{} \le \sqrt{\alpha}$, it requires $\bigomega{\degree{}}$ queries. Together, this gives a lower bound of $\Omega(\min(d, \alpha/d))$. 

For the case when the number of vertices $\vertexcount$ is unknown, we use a slightly different construction. Rather than varying the number of cliques, we vary the number of matching vertices. Here, the both the graphs contain exactly $\lbclqcnt$ cliques, but the number of matching vertices varies between $\vertexcount - \lbclqcnt$ and $2\vertexcount - \lbclqcnt$.

\noindent\textbf{Lower Bounds: \texorpdfstring{$\approxerror,\confidence$-Sensitive}{epsilon,delta-Sensitive} (\Cref{SubSec: EpsDelLB}).}
To capture the dependence on $\approxerror$ and $\delta$, we reduce from the Popcount Thresholding Problem (\ptp{}) \cite{DBLP:conf/approx/AssadiN22}. In this problem, given a binary string generated by i.i.d.\ Bernoulli trials, the goal is to distinguish whether the bias is $p(1-\approxerror)$ or $p(1+\approxerror)$. We encode the binary string into a graph by mapping each bit to a block: $1$-bits correspond to cliques and $0$-bits to independent sets. Under this construction, the average degree is proportional to the Hamming weight of the string.

To implement the reduction, we show how the graph queries can be simulated using queries to the binary string. We show \degreeq{}, \neighbourq{}, \edgeexistsq{}, and \fullnbrq{} can be simulated by $\bigo{1}$ queries to the string, while \randedgeq{} queries are simulated via rejection sampling. We then leverage known lower bounds for \ptp{} to establish our lower bounds.

\noindent\textbf{Lower Bounds: Power of \randedgeq{} (\Cref{SubSec:FeigeLB}).}
We construct two graph instances that differ only on a small subset of vertices. In one instance, this subset is isolated, while in the other it forms a complete bipartite graph with a larger component. The sizes of the vertex sets are chosen so that the bipartite subgraph contains $\bigomega{\approxerror \vertexcount \degree{}}$ edges.

The two instances remain indistinguishable as long as no vertex from, or edge incident to, this subset is observed. Under the stated query bounds, both \degreeq{} and \randedgeq{} queries sample this region with vanishing probability, and hence the two graphs remain indistinguishable.

\section{Preliminaries}\label{Section: Prelims}


In this section, we set up the preliminaries that will be relevant to our work. We start with the Nash-Williams Theorem, and two simple corollaries that will be relevant to our work.


\begin{lemma}[Nash-Williams Theorem~\cite{Nash-WilliamsArboricityTheorem}]
    Given a graph $\graph = \fbrac{\vertexset,\edgeset}$, and for a subgraph $H$ of $\graph$, denote $\vertexcount_H$, and $\edgecount_H$ to be its number of vertices and edges, respectively. Then, the arboricity $\arboricity$ of the graph $\graph$ satisfies:
    \begin{align*}
        \arboricity = \max_{H \subseteq \graph}\ceil{\frac{\edgecount_H}{\vertexcount_H-1}}
    \end{align*}
\end{lemma}

The next corollary provides an upper bound on the number of edges in a graph on $\vertexcount$ vertices, and arboricity $\arboricity$. The proof follows simply as $\graph$ is a valid subgraph of itself.

\begin{corollary}[Arboricity based bound on $\edgecount$]\label{Lemma: Arboricity Edge Bound}
    Given a graph on $\vertexcount$ vertices, and arboricity $\arboricity$, the number of edges, $\edgecount$ of the graph is bounded as $\edgecount \leq \vertexcount\arboricity$.
\end{corollary}

The next corollary provides upper bounds on the subgraphs, and directly follows from the theorem.

\begin{corollary}\label{Corollary: Subgraph Arboricity}
    Given a graph $\graph$ with arboricity $\arboricity_\graph$, the arboricity $\arboricity_H$ of any subgraph $H \subseteq \graph$ of $\graph$ satisfies $\arboricity_H \leq \arboricity_\graph$.
\end{corollary}

The next result establishes the existence of a good threshold depending only on the arboricity $\arboricity$ of the graph. 

\begin{lemma}\label{Lem: Arboricity based Good Threshold}
    Given a graph with arboricity $\arboricity$, a threshold $\threshold \geq p\arboricity$ ensures $\ledgecnt \geq  \edgecount\fbrac{1 - \frac{2}{p}}$. In particular, any threshold $\threshold \geq 8\arboricity$ satisfies $\ldens \geq \frac{3}{8}$.
\end{lemma}

\begin{proof}
    Given there are $\edgecount$ edges, the number of heavy vertices with respect to the threshold $\threshold \geq p\arboricity$ is at most $\nicefrac{2\edgecount}{p\arboricity}$. Notice that $\size{\edgeset_\heavyverts}$ is the size of the induced subgraph of the vertices $\heavyverts$. Let $\arboricity_\heavyverts$ denote the arboricity of this induced subgraph.
    
    By~\Cref{Corollary: Subgraph Arboricity}, we have $\arboricity_\heavyverts \leq \arboricity$. Then, by~\Cref{Lemma: Arboricity Edge Bound}, we have $\size{\edgeset(\heavyverts)} \leq \nicefrac{2\edgecount\arboricity_\heavyverts}{p\arboricity} \leq \nicefrac{2\edgecount}{p}$. Then, we have:
    \begin{align*}
        \ledgecnt = 2\size{\edgeset(\lightverts)} + \size{\edgeset(\heavyverts,\lightverts)} \geq \size{\edgeset(\lightverts)} + \size{\edgeset(\heavyverts,\lightverts)} = \edgecount - \size{\edgeset(\heavyverts)} \geq \edgecount\fbrac{1 - \frac{2}{p}}
    \end{align*}
\end{proof}

Some standard concentration inequalities are mentioned in~\Cref{App: ConcIneq} of the Appendix.

\section{The Algorithm}\label{Sec: Algorithm}

In this section, we present the algorithmic contribution of our work. We start by setting up the building blocks in~\Cref{SubSec: Building Blocks}. Next, we describe the algorithms that work with advice in~\Cref{SubSec: Est With Advice}, the first one with advice on both a good threshold, and the average degree, and the second one with advice only on a good threshold. Finally, in~\Cref{SubSec: Est w/o Advice}, we establish an algorithm that works without any advice. Some of the proofs are presented in \Cref{AppSec: Proofs for Upper Bounds}.

\subsection{The Building Blocks}\label{SubSec: Building Blocks}

In this section, we set up the building blocks of our algorithm: \ldensest{} that estimates the density of light edges $\ldens$, and \meanest{} that estimates the quantity $\nicefrac{\ledgecnt}{\vertexcount}$. The proofs are presented in \Cref{AppSec: Proof for Building Blocks}.

\subsubsection{\texorpdfstring{\ldensest{}}{CoinToss}}

We start with the algorithm \ldensest{} that takes as input a possible threshold $\threshold$, and the number of queries $\coinq$ it makes, and access to \degreeq{} and \randedgeq{} queries, and outputs $\estldens$ as an estimate of the quantity $\ldens$.

\begin{algorithm}[ht!]
    \caption{\ldensest{$\fbrac{\threshold,\qcolor{\coinq}}$}}\label{Alg: Light Density Estimator}
    \begin{algorithmic}[1]
        \Require Access to \degreeq{} and \randedgeq{} queries, a threshold $\threshold$, and  number of queries $\coinq$
        \For{$i \in [\qcolor{\coinq}]$}
            \State $(\altvertex,\vertex) \gets \randedgeq{}$ \Comment{Makes \qcolor{$\coinq$} \randedgeq{} queries}
            \State $w \gets \uniform\sbrac{\altvertex,\vertex}$
            \State $X_i \gets$ $1$ if $w \in \lightverts$ (i.e. $\degreeq{\fbrac{w}} \leq \threshold$), and $0$ otherwise(i.e. $\degreeq{\fbrac{w}} > \threshold$) \Comment{Makes \qcolor{$\coinq$} \degreeq{} queries}
        \EndFor
        \State \Return  $\estldens \gets \frac{1}{\coinq} \sum_{i\in[\coinq]} X_i$
    \end{algorithmic}
\end{algorithm}

The next lemma provides the performance guarantees of the~\Cref{Alg: Light Density Estimator}. The performance of the algorithm depends on the number of queries made and the quantity $\ldens$.

\begin{restatable}{lemma}{cointosslem}\label{Lem: DensEst Algo Guarantee}
    For the algorithm \ldensest{}, the following holds:
    \begin{enumerate}
        \item $\E\tbrac{\estldens} = \ldens$
        \item If $\coinq = \bigomega{\log\fbrac{1/\confidence}}$ and $\ldens < \frac{1}{4}$, i.e., the threshold $\threshold$ is not good, $\estldens \leq \frac{5}{16}$ with probability at least $1 - \confidence$.
        \item If $\coinq = \bigomega{\log\fbrac{1/\confidence}}$ and $\ldens \geq \frac{3}{8}$, the estimate satisfies $\estldens \geq \frac{5}{16}$ with probability at least $1 - \confidence$.
        \item If the threshold $\threshold$ is good, and $\coinq = \bigomega{\frac{\log\fbrac{1/\confidence}}{\approxerror^2}}$, $\estldens \in (1\pm\approxerror)\ldens$ with probability at least $1 - \confidence$.
    \end{enumerate}
\end{restatable}


\subsubsection{\texorpdfstring{\meanest{}}{MeanEst}}

Next, we describe the algorithm \meanest{} that is given a threshold $\threshold$, the number of queries $\degq$ it makes, and access to \degreeq{} query, and outputs $\dlest$ as an estimate of the quantity $\nicefrac{\ledgecnt}{\vertexcount}$. 

\begin{algorithm}[ht!]
    \caption{\meanest{$(\threshold,\qcolor{\degq})$}}\label{Alg: Light Degree Estimator}
    \begin{algorithmic}[1]
        \Require Access to \degreeq{} query, a good threshold $\threshold$, and number of queries \qcolor{$\degq$}
        \For{$i \in \qcolor{[\degq]}$}
            \State $d_i \gets \degreeq{}$ \Comment{Makes \qcolor{$\degq$} \degreeq{} queries}
            \State $w_i \gets d_i$ if $d_i \leq \threshold$, and $0$ otherwise
        \EndFor
        \State \Return $\dlest \gets \frac{1}{\degq} \sum_{i\in\degq} w_i$
    \end{algorithmic}
\end{algorithm}

The next lemma shows that if the number of queries $\degq$ made by \meanest{} is sufficiently high, then $\dlest$ is a $(1\pm\approxerror)$-multiplicative estimate to $\nicefrac{\ledgecnt}{\vertexcount}$.

\begin{restatable}{lemma}{meanestlem}\label{Lem: Low Degree Est Guarantee}
    If $\degq = \bigomega{\frac{\threshold}{\confidence\approxerror^2\degree{}}}$, the algorithm \meanest{} satisfies $\Pr\tbrac{\dlest \notin (1\pm\approxerror)\E[\dlest]} \leq \confidence$.
\end{restatable}

\subsection{Estimating with Advice}\label{SubSec: Est With Advice}

In this section, we present two algorithms, \alladv{} and \thradv{}. \alladv{} is given as advice a good threshold $\threshold$, a bound on the average degree $\degbnd$, and $\approxerror,\confidence$ to obtain an estimate $\degest$ that is a $(1\pm\approxerror)$-multiplicative approximation to the quantity $\degree{}$. \thradv{} then leverages \alladv{} to obtain an algorithm that does not require the advice $\degbnd$. The proofs are presented in \Cref{AppSec: Proof for Advice}.

\subsubsection{\texorpdfstring{\alladv{}}{AllAdvice}}

We begin with the algorithm \alladv{} that combines \ldensest{} and \meanest{} to get an algorithm to obtain an $(1\pm\approxerror)$-multiplicative estimate of the average degree with both advice $\threshold$, and $\degbnd$.

\begin{algorithm}    \caption{\alladv{$\fbrac{\threshold,\degbnd,\approxerror,\confidence}$}}\label{Algorithm: All Advice}
    \begin{algorithmic}[1]
        \Require A good threshold $\threshold$, $\degbnd$, $\approxerror \in \fbrac{0,1}$, and $\confidence \in \fbrac{0,\frac{1}{2}}$
        \State $\dlest \gets \meanest{\fbrac{\threshold, \qcolor{\degq} = \frac{\threshold}{\confidence\approxerror^2\degbnd}}}$ \Comment{Makes \qcolor{$\frac{\threshold}{\confidence\approxerror^2\degbnd}$} \degreeq{} queries.}
        \State $\estldens \gets \ldensest{\fbrac{\threshold, \qcolor{\coinq} = \frac{\constant\log\fbrac{1/\confidence}}{\approxerror^2}}}$ \Comment{Makes \qcolor{$\frac{\constant\log\fbrac{1/\confidence}}{\approxerror^2}$} \degreeq{} and \randedgeq{} queries.}
        \State \Return $\degest \gets \frac{\dlest}{\estldens}$
    \end{algorithmic}
\end{algorithm}

The following lemma provides the performance guarantee of the algorithm. The result combines the performance guarantees of \meanest{} (\Cref{Lem: DensEst Algo Guarantee}) and \ldensest{} (\Cref{Lem: Low Degree Est Guarantee}).
\begin{lemma}\label{Lemma: All Advice}
    \alladv{} makes $\degq = \bigo{\max\fbrac{\frac{\log\fbrac{1/\confidence}}{\approxerror^2},\frac{\threshold}{\confidence\approxerror^2\degbnd}}}$ \degreeq{}, and $\coinq = \bigo{\frac{\log\fbrac{1/\confidence}}{\approxerror^2}}$ \randedgeq{} queries, and if $\degbnd \leq 16\degree{}$, returns $\degest \in (1\pm\approxerror)\degree{}$ with probability at least $1 - \confidence$. Furthermore, for all values of $\degbnd$, the estimator satisfies $\E[\degest] \leq 8\degree{}$ with probability at least $1 - \confidence$.
\end{lemma}

\begin{proof}
    The query bounds follow directly from the algorithm, as highlighted in the comments of the pseudocode. Then, by~\Cref{Lem: DensEst Algo Guarantee}, we have $\estldens \in \fbrac{1\pm\nicefrac{\approxerror}{2}}\ldens$, and by~\Cref{Lem: Low Degree Est Guarantee}, we have $\dlest \in \fbrac{1\pm\nicefrac{\approxerror}{2}}\nicefrac{\ledgecnt}{\vertexcount}$ with probability at least $1- \nicefrac{\confidence}{2}$. Combining through union bound, we have $\degest = \nicefrac{\dlest}{\estldens} \in \fbrac{1\pm\approxerror}\nicefrac{\ledgecnt}{\vertexcount\ldens} = (1\pm\approxerror)\degree{}$ with probability at least $1 - \confidence$.

    For the expectation, observe that $\degest = \nicefrac{\dlest}{\estldens}$. By \Cref{Lemma: Arboricity Edge Bound}~and~\Cref{Lemma: Low Degree Estimator Expectation}, for any good threshold, we have $\E[\dlest] = \nicefrac{\ledgecnt}{\vertexcount} \leq \degree{}$. By~\Cref{Lem: DensEst Algo Guarantee}, for any good threshold, we have $\estldens \geq \nicefrac{\ldens}{2} \geq \nicefrac{1}{8}$ with probability at least $1 - \confidence$. Combining these two bounds completes the proof.
\end{proof}


\subsubsection{\texorpdfstring{\thradv{}}{ThresholdAdvice}}

Now, we present \thradv{}, which does not require the advice $\degbnd$ on the average degree of the graph. The algorithm leverages the \alladv{} algorithm to search for an appropriate bound $\degbnd$ on the average degree. This form of parameter search is widely used in the property testing literature, see for example Eden et al.~\cite{EdenLRS/SIAMJComp/2017/SublinearTriangle,EdenRS/SODA/2020/kCliquesArboricity}.


\begin{algorithm}[ht]
    \caption{\thradv{$\fbrac{\threshold,\approxerror,\confidence}$}}\label{Algorithm: Thr Advice}
    \begin{algorithmic}[1]
        \Require A good threshold $\threshold$, $\approxerror \in \fbrac{0,1}$, and $\confidence \in \fbrac{0,\frac{1}{2}}$
        \For{$i \in [\log\threshold]$}\label{Outerloop}\Comment{Outer Loop}
            \State $\degbnd_i \gets \frac{\threshold}{2^i}$
            \For{$j \in [\ceil{\constant \log (\frac{\log \threshold}{\confidence})}]$}\label{Inner Loop}\Comment{Inner Loop}
                \State $\degest_j \gets \alladv{\fbrac{\threshold,\degbnd_i,\approxerror,\frac{\confidence}{\constant\log^2{\threshold}}}}$ \Comment{Makes \qcolor{$\frac{\constant\threshold\log^2\threshold}{\confidence\approxerror^2\degbnd_i}$} \degreeq{} and \qcolor{$\frac{\constant\log\fbrac{\log\threshold/\confidence}}{\approxerror^2}$} \randedgeq{} queries.}
            \EndFor
            \State $\degree{min} \gets \min_{j} \degest_j$
            \If{$\degree{min} \geq \degbnd_i$}\Comment{Valid $\degbnd$}\label{Valid Degbnd}
                \State \Return $\degest \gets \alladv{\fbrac{\threshold,\degbnd_i,\approxerror,\frac{\confidence}{2}}}$ \Comment{Makes \qcolor{$\frac{\constant\threshold}{\confidence\approxerror^2\degbnd_i}$} \degreeq{} and \qcolor{$\frac{\constant\log\fbrac{1/\confidence}}{\approxerror^2}$} \randedgeq{} queries.}
            \EndIf
        \EndFor
    \end{algorithmic}
\end{algorithm}

Now, we establish the performance guarantees of the \thradv{} algorithm.

\begin{restatable}{lemma}{thresadvlem}\label{Lem: Thr Advice Guarantee}
    \thradv{} returns $\degest \in (1\pm\approxerror)\degree{}$ with probability at least $1 - \confidence$, and makes $\bigo{\frac{\threshold\log^4\threshold}{\confidence\approxerror^2\degree{}}}$ \degreeq{}, and $\bigo{\log^2\threshold\frac{\log(\log\threshold/\confidence)}{\approxerror^2}}$ \randedgeq{} queries in expectation.
\end{restatable}

\subsection{Estimating without Advice - \texorpdfstring{\noadv{}}{NoAdvice}}\label{SubSec: Est w/o Advice}

Now we present the algorithm \noadv{} that can estimate the average degree without any advice. The idea here is that we do not need to actually know $\arboricity$ for our algorithm to succeed. Any $\threshold$ that satisfies the good threshold property will be sufficient, and we can verify the goodness of a threshold using the \ldensest{} algorithm. On the other hand, Lemmas~\ref{Lem: Arboricity based Good Threshold}~and~\ref{Lem: DensEst Algo Guarantee} show that any value larger than $8\arboricity$ will ensure $\estldens \geq \nicefrac{5}{8}$. We combine these two observations to search for a good threshold and bound the number of queries made by the algorithm. The proofs are presented in \Cref{AppSec: Proof for NoAdv}.


\begin{algorithm}
    \caption{\noadv{$\fbrac{\approxerror,\confidence}$}}\label{Algorithm: No Advice}
    \begin{algorithmic}[1]
        \Require $\approxerror \in \fbrac{0,1}$, and $\confidence \in \fbrac{0,\frac{1}{2}}$
        \State $i \gets 0$
        \While{\texttt{True}}
            \State $\threshold_i \gets 2^i$
            \State $\estldens_i \gets \ldensest{\fbrac{\threshold_i,\constant\log\frac{4\threshold_i}{\confidence}}}$ \Comment{Makes \qcolor{$\constant\log\frac{4\threshold_i}{\confidence}$} \randedgeq{} queries}
            \If{$\estldens_i \geq \frac{5}{16}$}\label{ThresDetect}\Comment{Getting good threshold}
                \State \Return $\degest \gets \thradv{\fbrac{\threshold_i,\approxerror,\frac{\confidence}{2}}}$\Comment{Makes  $\qcolor{\frac{\constant\threshold_i\log^3\threshold_i}{\confidence\approxerror^2\degree{}}}$ \degreeq{} and $\qcolor{\log^2\threshold_i\frac{\constant\log(\log\threshold_i/\confidence)}{\approxerror^2}}$ \randedgeq{} queries}
            \EndIf
            \State $i \gets i+1$
        \EndWhile
    \end{algorithmic}
\end{algorithm}

The next result provides the performance guarantees of \noadv{}. Here, we combine the performance guarantees of \ldensest{} (\Cref{Lem: DensEst Algo Guarantee}), and \thradv{} (\Cref{Lem: Thr Advice Guarantee}) along with the role of arboricity as a good threshold (\Cref{Lem: Arboricity based Good Threshold}) to obtain our result.

\begin{restatable}[Estimating Without Advice]{lemma}{noadvlem}\label{Lemma: No Advice Guarantee}
    \noadv{} makes $\bigot{\frac{\arboricity}{\confidence\approxerror^2\degree{}}}$ \degreeq{} queries, and $\bigot{\frac{\log\fbrac{1/\confidence}}{\approxerror^2}}$ \randedgeq{} queries, and returns $\degest \in (1\pm\approxerror)\degree{}$ with probability at least $1 - \confidence$.
\end{restatable}

\subsection{Final Bounds}\label{SubSec: Final}

In this section, we state the main upper bounds of our work. First, we establish an upper bound for the case of general graphs by fixing $\confidence = \nicefrac{2}{3}$.

\UBG*


Furthermore, we can extend this result using an algorithm of Beretta and T\v{e}tek~\cite{BerettaTetek/TALG/2024/BetterSumEstimationViaWeightedSampling} to obtain the following result. The details are presented in \Cref{AppSec: Proof for Final}.

\UBNI*

\section{Lower Bounds}\label{Sec: Lower Bounds}

In this section, we establish the lower bound of our work. \Cref{SubSec: GeneralLB} presents the graph parameter based lower bounds (\Cref{Thm: Lower Bound,Thm: Lower Bound Unknown N}), for both the case of known and unknown $\vertexcount$. \Cref{SubSec: EpsDelLB} presents $\fbrac{\approxerror,\confidence}$-sensitive lower bound (\Cref{Theorem: Lower Bound Epsilon Delta}). Finally, \Cref{SubSec:FeigeLB} presents the lower bound regarding the requirement of \randedgeq{} queries (\Cref{Thm: LBFeige}).

\subsection{Lower Bounds: Graph Parameters}\label{SubSec: GeneralLB}

We start with the \Cref{Thm: Lower Bound}, where the algorithm is allowed to have the knowledge of $\vertexcount$ and $\arboricity$. We outline the graph construction in \Cref{fig: Lower Bound Construction}. The two families of graphs are constructed by varying $\lbclqcnt$, the numbers of cliques in the graph. The details are provided in \Cref{AppSec: Generic LB Proofs}.

\begin{wrapfigure}{R}{0.6\textwidth}
    \centering

\begin{tikzpicture}[
  vertex/.style={circle, draw, fill=blue!15, inner sep=1.5pt},
  matchingnode/.style={circle, draw, fill=red!15, inner sep=2pt},
  matchingedge/.style={thick, red},
  cliqueedge/.style={thin, gray!70},
  scale=0.7
]

\newcommand{\drawclique}[1]{
  \foreach \i in {1,...,5}
    \node[vertex] (v\i#1) at ($(#1)+(72*\i:0.7)$) {};
  \foreach \i in {1,...,5}{
    \foreach \j in {\i,...,5}{
      \ifnum\i<\j
        \draw[cliqueedge] (v\i#1) -- (v\j#1);
      \fi
    }
  }
}

\coordinate (C1) at (0,0);
\coordinate (C2) at (0,-2);
\coordinate (C3) at (0,-5); 

\drawclique{C1}
\drawclique{C2}
\drawclique{C3}

\draw[dotted, thick] (0,-3) -- (0,-4.2);

\draw[decorate,decoration={brace,amplitude=8pt}] 
  (-1.6,-6) -- (-1.6,0.8);
\node[rotate=90, anchor=south] at (-1.9,-2.6)
  {$\lbclqcnt$ cliques of size $\arboricity$};

\begin{scope}[xshift=5cm]

  \node[matchingnode] (m1a) at (0,0) {};
  \node[matchingnode] (m1b) at (1,0) {};
  \draw[matchingedge] (m1a) -- (m1b);

  \node[matchingnode] (m2a) at (0,-1.25) {};
  \node[matchingnode] (m2b) at (1,-1.25) {};
  \draw[matchingedge] (m2a) -- (m2b);

  \node[matchingnode] (m3a) at (0,-2.5) {};
  \node[matchingnode] (m3b) at (1,-2.5) {};
  \draw[matchingedge] (m3a) -- (m3b);

  \draw[dotted, thick] (0.5,-3.4) -- (0.5,-4.2);

  \node[matchingnode] (m4a) at (0,-5) {};
  \node[matchingnode] (m4b) at (1,-5) {};
  \draw[matchingedge] (m4a) -- (m4b);

  \draw[decorate,decoration={brace,amplitude=8pt}]
    (2.5,0.8) -- (2.5,-6);
  \node[rotate=90, anchor=north] at (2.8,-2.6)
        {matching on $\fbrac{\vertexcount - \lbclqcnt\arboricity}$ vertices};

\end{scope}

\end{tikzpicture}
    
    \caption{The construction for \Cref{Thm: Lower Bound - High Degree,Thm: Lower Bound - Low Degree,Thm: Lower Bound})}
    \label{fig: Lower Bound Construction}
\end{wrapfigure}


\LB*

Next, we establish lower bounds for the case when $\vertexcount$ is not known to the algorithm. Note that the result holds even when the algorithm is allowed the knowledge of arboricity of the graph.

\LBUnknownN*

\begin{proof}
    We consider graph constructions similar to that used in \Cref{Thm: Lower Bound - Low Degree,Thm: Lower Bound - High Degree}. Here, the two graphs we consider differ in the number of vertices. Both the graphs contain $\lbclqcnt = \frac{\vertexcount\degree{}}{\arboricity^2}$ cliques of size $\arboricity$, and the remaining vertices form a matching. One graph contains $\vertexcount$ vertices, and the other contains $2\vertexcount$ vertices. In this case also, we have $\size{\lbclqv} = \lbclqcnt\arboricity$, $\size{\lbmatv} = \bigtheta{\vertexcount - \lbclqcnt\arboricity}$, $\size{\lbclqe} = \bigtheta{\lbclqcnt\arboricity^2}$, and $\size{\lbmate} = \bigtheta{\vertexcount - \lbclqcnt\arboricity}$.

    We now want to show that an algorithm that makes $\smallo{\min\fbrac{\degree{},\frac{\arboricity}{\degree{}}}}$ queries can not distinguish between these two cases. By a similar argument as in \Cref{Thm: Lower Bound - Low Degree,Thm: Lower Bound - High Degree}, the samples received by \degreeq{} and \randedgeq{} queries lies in $\lbmatv$ and $\lbclqv$, respectively and are distinct. Consequently, \neighbourq{}, \edgeexistsq{}, and \fullnbrq{} can not distinguish between the two cases. Hence, the graphs remain indistinguishable.
\end{proof}

\subsection{Lower Bounds: \texorpdfstring{$\approxerror,\confidence$-Sensitive}{epsilon,delta-Sensitive}}\label{SubSec: EpsDelLB}

In this section, we establish lower bounds on estimating average degree $\degree{}$ that depends on $\approxerror$, and $\confidence$ as well as the graph parameters. We obtain the lower bound by reducing the the popcount thresholding problem (henceforward referred to as \ptp{}) (\Cref{Def: PTP}) to edge estimation. The \ptp{} problem asks to determine whether a binary string is generated from one of the two distributions that vary in the probability of outputting $1$. The encoding is as described in \Cref{Fig: LBEpsDel}, each bit having value $1$ correspond to a clique, and each bit having value $0$ correspond to an independent set. The details are presented in \Cref{AppLBSec: EpsDel Proof}.

\begin{figure}[ht]
    \centering

\begin{tikzpicture}[
  vertex/.style={circle, draw, fill=blue!15, inner sep=1.5pt},
  cliqueedge/.style={thin, gray!70},
  scale=0.8
]

\newcommand{\drawclique}[1]{
  \foreach \i in {1,...,5}
    \node[vertex] (v\i#1) at ($(#1)+(72*\i:0.7)$) {};
  \foreach \i in {1,...,5}{
    \foreach \j in {\i,...,5}{
      \ifnum\i<\j
        \draw[cliqueedge] (v\i#1) -- (v\j#1);
      \fi
    }
  }
}

\newcommand{\drawindep}[1]{
  \foreach \i in {1,...,5}
    \node[vertex] (v\i#1) at ($(#1)+(72*\i:0.7)$) {};
}


\node at (1.5, 2.5) {\Large $x_1 = 1$};
\node at (4.5, 2.5) {\Large $x_2 = 0$};
\node at (7.5, 2.5) {\Large $x_3 = 1$};
\node at (10, 2.5) {\Large $\dots$};
\node at (12.5, 2.5) {\Large $x_\stringlength = 0$};

\draw[dashed, thick] (1.5,2.15) -- (1.5,1.35);
\draw[dashed, thick] (4.5,2.15) -- (4.5,1.35);
\draw[dashed, thick] (7.5,2.15) -- (7.5,1.35);
\draw[dashed, thick] (12.5,2.15) -- (12.5,1.35);

\coordinate (V1) at (1.5, 0);
\coordinate (V2) at (4.5, 0);
\coordinate (V3) at (7.5, 0);
\coordinate (Vs) at (12.5, 0);

\drawclique{V1}
\drawindep{V2}
\drawclique{V3}
\drawindep{Vs}

\node at (1.5, -1.5) {$V_1$};
\node at (4.5, -1.5) {$V_2$};
\node at (7.5, -1.5) {$V_3$};
\node at (12.5, -1.5) {$V_\stringlength$};

\node[text width=2.5cm, align=center] at (1.5, -2.6) {Clique\\(size $\alpha$)};
\node[text width=2.5cm, align=center] at (4.5, -2.6) {Independent Set\\(size $\alpha$)};
\node[text width=2.5cm, align=center] at (7.5, -2.6) {Clique\\(size $\alpha$)};
\node[text width=2.5cm, align=center] at (12.5, -2.6) {Independent Set\\(size $\alpha$)};

\draw[dashed, gray] (0.2, 1.2) rectangle (2.8, -1.8);
\draw[dashed, gray] (3.2, 1.2) rectangle (5.8, -1.8);
\draw[dashed, gray] (6.2, 1.2) rectangle (8.8, -1.8);
\draw[dashed, gray] (11.2, 1.2) rectangle (13.8, -1.8);

\end{tikzpicture}

\caption{The construction for \Cref{Theorem: Lower Bound Epsilon Delta}, mapping the \ptp{} string $x \in \sbrac{0,1}^\stringlength$ to the graph $G_x$.}
\label{Fig: LBEpsDel}
\end{figure}

\LBEpsDel*

\subsection{Lower Bounds: Power of \texorpdfstring{\randedgeq{}}{RandEdge}}\label{SubSec:FeigeLB}

Here, we establish the requirement of \randedgeq{} queries to achieve sublinear time algorithms using \degreeq{} and \randedgeq{} queries. The graph construction is outlined in \Cref{Fig: LBFeige}.

\begin{figure}[ht]

    \centering

\begin{tikzpicture}[
  vertex/.style={circle, draw, fill=blue!15, inner sep=1pt},
  heavynode/.style={circle, draw, fill=green!20, inner sep=1pt},
  cliqueedge/.style={thin, gray!70},
  regedge/.style={thick, red!70},
  bipedge/.style={thick, dashed},
  scale=0.7
]

\newcommand{\drawclique}[1]{
  \foreach \i in {1,...,5}
    \node[vertex] (v\i#1) at ($(#1)+(72*\i:0.7)$) {};
  \foreach \i in {1,...,5}{
    \foreach \j in {\i,...,5}{
      \ifnum\i<\j
        \draw[cliqueedge] (v\i#1) -- (v\j#1);
      \fi
    }
  }
}

\newcommand{\drawregular}[1]{
  \foreach \i in {1,...,4}
    \node[heavynode] (r\i#1) at ($(#1)+(90*\i:0.5)$) {};
  \draw[regedge] (r1#1) -- (r2#1);
  \draw[regedge] (r2#1) -- (r3#1);
  \draw[regedge] (r3#1) -- (r4#1);
  \draw[regedge] (r4#1) -- (r1#1);
}

\node at (-4,1.5) {$G_1$};

\node[vertex] (i1) at (-6,0) {};
\node[vertex] (i2) at (-6,-2) {};
\draw[dotted, thick] (-6,-3) -- (-6,-4.2);
\node[vertex] (i3) at (-6,-5) {};

\coordinate (R1) at (-4.5,0);
\coordinate (R2) at (-4.5,-2);
\coordinate (R3) at (-4.5,-5);

\drawregular{R1}
\drawregular{R2}
\draw[dotted, thick] (-4.5,-3) -- (-4.5,-4.2);
\drawregular{R3}

\coordinate (C1) at (-2,0);
\coordinate (C2) at (-2,-2);
\coordinate (C3) at (-2,-5);

\drawclique{C1}
\drawclique{C2}
\draw[dotted, thick] (-2,-3) -- (-2,-4.2);
\drawclique{C3}

\node at (-6,-6.5) {$V_I$};
\node at (-4.5,-6.5) {$V_R$};
\node at (-2,-6.5) {$V_C$};

\node[align=center] at (-3.5,-8.2) {No edges between $V_I$ and $V_R$};

\begin{scope}[xshift=9cm]

\node at (-4,1.5) {$G_2$};

\node[vertex] (i1r) at (-6,0) {};
\node[vertex] (i2r) at (-6,-2) {};
\draw[dotted, thick] (-6,-3) -- (-6,-4.2);
\node[vertex] (i3r) at (-6,-5) {};

\coordinate (R1r) at (-4.5,0);
\coordinate (R2r) at (-4.5,-2);
\coordinate (R3r) at (-4.5,-5);

\drawregular{R1r}
\drawregular{R2r}
\draw[dotted, thick] (-4.5,-3) -- (-4.5,-4.2);
\drawregular{R3r}

\coordinate (C1r) at (-2,0);
\coordinate (C2r) at (-2,-2);
\coordinate (C3r) at (-2,-5);

\drawclique{C1r}
\drawclique{C2r}
\draw[dotted, thick] (-2,-3) -- (-2,-4.2);
\drawclique{C3r}


\draw[bipedge] (i1r) -- (r1R1r);
\draw[bipedge] (i1r) -- (r3R3r);

\draw[bipedge] (i2r) -- (r1R1r);
\draw[bipedge] (i2r) -- (r3R3r); 

\draw[bipedge] (i3r) -- (r1R1r);
\draw[bipedge] (i3r) -- (r3R3r); 

\node at (-6,-6.5) {$V_I$};
\node at (-4.5,-6.5) {$V_R$};
\node at (-2,-6.5) {$V_C$};

\node[align=center] at (-3.5,-8.2) {$V_I \times V_R$ is complete bipartite};

\end{scope}

\end{tikzpicture}

\caption{The construction for \Cref{Thm: LBFeige}}\label{Fig: LBFeige}
\end{figure}

\LBFeige*

\begin{proof}
    We consider two graph instances, $\graph_1$ and $\graph_2$. Both the graphs contain four sets of vertices, $\vertexset_I,\vertexset_C,\vertexset_R$, and $\vertexset_L$. For both the graphs, we have $\size{\vertexset_I} = 2\approxerror\degree{}$, $\size{\vertexset_C} = \frac{\vertexcount\degree{}}{\arboricity}$, $\size{\vertexset_R} = \frac{\vertexcount}{2}$, and $\vertexset_L$ contains the remaining vertices. Note that for our choices of $\vertexcount,\degree{},\arboricity$, and $\approxerror$, we have $2\approxerror\degree{} + \frac{\vertexcount\degree{}}{\arboricity} \leq \frac{\vertexcount}{2}$.
   
    We now describe the set of edges. For both the graphs, $\vertexset_C$ contains $\frac{\vertexcount\degree{}}{2\arboricity^2}$ cliques of size $2\arboricity$, and $\vertexset_L$ is an independent set. In $\graph_1$, $\vertexset_R$ forms a $\degree{}$-regular  graph, and $\vertexset_I$ is an independent set. In $\graph_2$,  $\vertexset_R$ forms a $\fbrac{\degree{}-\approxerror\degree{}}$-regular graph, and $\vertexset_I$ and $\vertexset_R$ forms a complete bi-partite graph.

    Observe that, the cliques of size $2\arboricity$ ensure that the arboricity is at least $\arboricity$. Also, given $\degree{} \leq \frac{\arboricity}{4}$, the arboricity for the induced subgraph on $\vertexset_I\cup\vertexset_R$ is always less than $\arboricity$. Hence, the arboricity of both the graphs is exactly $\arboricity$. The number of edges in the graphs can be at most $
    2\arboricity^2\frac{\vertexcount\degree{}}{2\arboricity^2}+\frac{\vertexcount\degree{}}{2} = \frac{3\vertexcount\degree{}}{2}
    $
    
    Here, the first term corresponds to the number of edges within $\vertexset_C$, and the second term corresponds to the number of edges incident on $\vertexset_R$. Furthermore, the total number of edges is at least
    $
    2\arboricity^2\frac{\vertexcount\degree{}}{2\arboricity^2}+\frac{\vertexcount\degree{}}{4} = \frac{5\vertexcount\degree{}}{4}
    $
    Furthermore, for the graph $\graph_2$, the degrees of all vertices in $\vertexset_C,\vertexset_R$, and $\vertexset_L$ remains same while the degree of all $\approxerror\degree{}$ vertices in $\vertexset_I$ increases from $0$ to $\frac{\vertexcount}{2}$. Hence, the average degree of $\graph_2$ increases by $\approxerror\degree{}$. Thus, any algorithm that obtains a $(1\pm\approxerror)$-approximation of the average degree should be able to distinguish between these two instances. 

    We now show that an algorithm making $\smallo{\frac{1}{\approxerror}}$ \randedgeq{} query and $\smallo{\frac{\vertexcount}{\approxerror\degree{}}}$ \degreeq{} query can not distinguish between $\graph_1$, and $\graph_2$. Note that, to distinguish between the two cases, the algorithm has to either obtain the degree of a vertex in $\vertexset_I$, or an edge incident on $\vertexset_I$. The probability that a $\randedgeq{}$ query hits any of the edges incident on $\vertexset_I$ is at most $\bigo{\approxerror}$. Hence, by union bound, with probability at least $1 - \smallo{1}$, none of the \randedgeq{} queries hit an edge incident on $\vertexset_I$. We now consider the \degreeq{} queries. The probability that any \degreeq{} query hits a vertex in $\vertexset_I$ is at most $\frac{\approxerror\degree{}}{\vertexcount}$.  Hence, by union bound, with probability at least $1 - \smallo{1}$, none of the \degreeq{} queries hit a vertex in $\vertexset_I$. Thus, the algorithm can not distinguish between the two cases.
\end{proof}

\clearpage
\phantomsection
\addcontentsline{toc}{section}{\refname}
\bibliography{refs} 

\appendix
\section*{Appendix}

\section{Further Discussion on the Results}\label{SubSec: Discussion}

In this section, we discuss the implications of our results, beginning with an overview of the main consequences of our bounds.

    \subsection{Comparison with Weighted Sampling Query Access:}
    Our results show that the improvement previously achieved using additional \neighbourq{} queries can, in fact, be obtained solely by leveraging the underlying graph structure, without requiring any extra query access. This advances the line of work on weighted sampling access models, where Beretta and T\v{e}tek~\cite{BerettaTetek/TALG/2024/BetterSumEstimationViaWeightedSampling} established a $\bigomega{\sqrt{\vertexcount}}$ lower bound for the more general problem of sum estimation, and Beretta et al.~\cite{BerettaCS/ArXiv/2025/FasterEdgeEstimation} later overcame this bound by augmenting the model with \neighbourq{} queries. In contrast, our bounds demonstrate that comparable improvements arise naturally from exploiting structural properties of the graph itself.
    \subsection{Parameter-oblivious algorithms:}
     Obtaining size-oblivious query based algorithms is an interesting problem across query models~\cite{Goldreich/OpenProbsGraphPT}. Our algorithm is an effort in that direction as it operates without prior knowledge of the number of vertices $\vertexcount$ or the arboricity $\arboricity$. Unlike previous algorithms designed for unknown $\vertexcount$~\cite{motwani07,BerettaTetek/TALG/2024/BetterSumEstimationViaWeightedSampling,BerettaCS/ArXiv/2025/FasterEdgeEstimation}, it does not rely on searching for an appropriate guess of $\vertexcount$, thereby avoiding additional $\poly{\log\vertexcount}$ factors in its query complexity. Moreover, although $\arboricity$ offers a valid upper bound during the search phase, our algorithm can terminate well before reaching the value of $\arboricity$.
    \subsection{Practical relevance:}
    \randedgeq{} queries have been widely studied in the data mining community in the context of edge sampling~\cite{RibeiroTowsley/CDC/2012/DegreeDistributionGraphSampling,AhmedNevilleKompella/TKDD/2014/NetworkSamplingStaticToStreaming,LeskovecFaloutsos/KDD/2006/SamplingFromLargeGraphs}. These queries are easy to implement in practice since many real-world graphs are stored as edge lists, allowing random access to edges directly~\cite{snapnets,NetworkRepo-aaai15}. Since the work of Aliakbarpour et al.~\cite{Aliakbarpour/Algorithmica/2018/StarSubgraphEdgeSampling}, \randedgeq{} queries have also been studied extensively for sublinear-time graph algorithms~\cite{assadi2018simple,EdenRosenbaum/Approx/2018/LowerBoundGraphCommunication,TetekThorup/STOC/2022/EdgeSamplingFullNbrhood,bishnu2025arboricityrandomedgequeries,BerettaCS/ArXiv/2025/FasterEdgeEstimation,EdenRV/ArXiV/2025/TestableAlgoTrianglEdges}, often yielding simple and efficient procedures. In this context, our algorithm stands out for its simplicity and ease of implementation.

    \subsection{Regarding the questions of~Beretta et al.\texorpdfstring{~\cite{BerettaCS/ArXiv/2025/FasterEdgeEstimation}}{}:} Beretta et al.~\cite{BerettaCS/ArXiv/2025/FasterEdgeEstimation} raised two explicit questions for future work. Beretta et al. [\cite{BerettaCS/ArXiv/2025/FasterEdgeEstimation}, Question 1.7]  ask whether the $\polylogneps$ terms in their bound can be improved, and in particular avoid the binning technique involved in their work. Our algorithm does not employ any binning technique, and does not have any $\poly{\log\vertexcount}$ factors in the query complexity. This partially resolves the first question of~Beretta et al.~\cite{BerettaCS/ArXiv/2025/FasterEdgeEstimation}. Secondly, Beretta et al. [\cite{BerettaCS/ArXiv/2025/FasterEdgeEstimation}, Question 1.8] asked whether their bounds can be parametrized by $\arboricity$. Our upper bound provides an algorithm that uses only \degreeq{} and \randedgeq{} to achieve this, avoiding the additional access to \neighbourq{} in their algorithm. On the other hand, our lower bounds establish that any algorithm that works for all ranges of $\arboricity$, even with access to all the queries they considered, must make $\bigomega{\nicefrac{\arboricity}{\degree{}}}$ queries. This resolves the second question of~Beretta et al.~\cite{BerettaCS/ArXiv/2025/FasterEdgeEstimation}. 




\section{Proofs for \texorpdfstring{\Cref{Sec: Algorithm}}{Section 5}}\label{AppSec: Proofs for Upper Bounds}

\subsection{Proofs for \texorpdfstring{\Cref{SubSec: Building Blocks}}{Section 5.1}}\label{AppSec: Proof for Building Blocks}

\cointosslem*

\begin{proof}
    First, we note that the $X_i$-s are defined in~\ldensest{} are i.i.d. Bernoulli random variables with $p = \ldens$.
    \begin{align*}
        \Pr[X_i = 1] = \sum_{\lightvert \in \lightverts} \sum_{i \in [\degree{\lightvert}]} \frac{1}{2\edgecount} = \sum_{\lightvert \in \lightverts} \frac{\degree{\lightvert}}{2\edgecount} = \frac{\ledgecnt}{2\edgecount} = \ldens
    \end{align*}
    The result of part (1) follows directly by taking expectation. We obtain the parts (2) and (3) of the claim through an additive Chernoff bound with additive error $\nicefrac{1}{16}$. For the part (4), as the threshold is good, we have $\ldens \geq \nicefrac{1}{4}$. Hence, by multiplicative Chernoff bound, the claim holds.
\end{proof}

Next, we want to show that $\dlest$ gives a good approximation to $\nicefrac{\ledgecnt}{\vertexcount}$ if the number of \degreeq{} queries $\degq$ is sufficiently high. First, we state the expectation of $\dlest$.

\begin{restatable}{lemma}{meanestbias}\label{Lemma: Low Degree Estimator Expectation}
    In the algorithm \meanest{}, we have $\E[\dlest] = \frac{\ledgecnt}{\vertexcount}$.
\end{restatable}

\begin{proof}
    Note that only  vertices in $\lightverts$ contribute to the estimate. Hence,
    \begin{align*}
        \E[\dlest] =  \E\tbrac{\frac{1}{\degq} \sum_{i \in [\degq]} w_i} = \frac{1}{\degq} \sum_{i \in [\degq]} \E[w_i] = \E[w_i] = \sum_{\lightvert \in \lightverts} \frac{1}{\vertexcount} \degree{\lightvert} = \frac{\ledgecnt}{\vertexcount}
    \end{align*}
\end{proof}

Next, we bound the variance of the estimate $\dlest$.

\begin{restatable}{lemma}{meanestvar}\label{Lem: Low Deg Var Bound}
    In the algorithm \meanest{}, we have $\Var[\dlest] \leq \frac{\threshold}{\degq} \E[\dlest]$.
\end{restatable}

\begin{proof}
    First observe that $w_i$ and $w_j$ are independent $\forall i,j \in [\degq], i\neq j$. Hence,
    \begin{align*}
        \Var[\dlest] = \Var\tbrac{\frac{1}{\degq}\sum_{i \in [\degq]} w_i} = \frac{1}{\degq^2} \Var\tbrac{\sum_{i \in [\degq]} w_i} = \frac{1}{\degq^2} \sum_{i \in [\degq]} \Var[w_{i}]
    \end{align*}
    Here, the last equality is due to the fact that $\Cov(w_i,w_j) = 0$ for all $i,j \in [\degq]$. Now, we focus on the individual $w_i$s,
    \begin{align*}
        \Var[w_i] \leq \E[w_i^2] \leq \threshold \E[w_i] = \threshold \E[\dlest]
    \end{align*}
    Here, the second inequality is due to the fact that $\threshold \geq w_i$ for all $i \in [\degq]$. Combining the results above, we have:
    \begin{align*}
        \Var\tbrac{\frac{1}{\degq}\sum_{i \in [\degq]} w_i} \leq \frac{\threshold}{\degq} \E[\dlest] 
    \end{align*}
\end{proof}

\meanestlem*

\begin{proof}
    By Chebyshev's inequality, we have:
    \begin{align*}
        \Pr\tbrac{\dlest \notin (1\pm\approxerror)\E[\dlest]} \leq \frac{\Var[\dlest]}{\approxerror^2\E^2[\dlest]} \leq \frac{\threshold}{\approxerror^2 \degq \E[\dlest]} \leq \frac{4\threshold}{\approxerror^2 \degq \degree{}} \leq \confidence.
    \end{align*}
    Here, the third inequality follows from the fact that the given threshold $\threshold$ is good.
\end{proof}

\subsection{Proofs for \texorpdfstring{\Cref{SubSec: Est With Advice}}{Section 5.2}}\label{AppSec: Proof for Advice}

\thresadvlem*

\begin{proof}
    We start with the guarantees of the estimate $\degest$ produced by the algorithm. First, we consider any value of $\degbnd$ such that $\degbnd \geq 16\degree{}$. The bound on $\E[\degest]$ of~\Cref{Lemma: All Advice} holds throughout all the runs with probability at least $1 - \nicefrac{\confidence}{3}$ by union bound. Conditioning on this event, a Markov's inequality argument based on~\Cref{Lemma: All Advice} shows that for any $j$, we have $\Pr\tbrac{\degest_j \geq \degbnd} \leq \nicefrac{1}{2}$, for any $\degbnd \geq 8\degree{}$.  The probability that $\degree{min}$, the minimum over $\constant \log \fbrac{\log \threshold/\confidence}$ such estimates, is smaller than $\degbnd$ is at most $\nicefrac{\confidence^\constant}{\log^\constant\threshold}$. There are at most $\log\threshold$ such values that appear in this algorithm due to the outer loop (Line~\ref{Outerloop}). Hence, the probability that the algorithm returns $\degest$ by satisfying the valid $\degbnd$ criteria at Line~\ref{Valid Degbnd} for some $\degbnd \geq 16\degree{}$ is at most $\nicefrac{\confidence}{3}$. 
    
    Alternatively, if a valid $\degbnd$, i.e. $\degbnd \leq 16\degree{}$ satisfies the condition, then by~\Cref{Lemma: All Advice}, it returns $\degest \in \fbrac{1\pm\approxerror}\degree{}$ with probability at least $1-\nicefrac{\confidence}{3}$. Combining, we obtain that the algorithm returns $\degest \in \fbrac{1\pm\approxerror}\degree{}$ with probability at least $1-\confidence$.

    Now, we bound the number of queries made by the algorithm. Note that for all values of $\degbnd$ such that  $\degbnd > \nicefrac{\degree{}}{8}$, the query complexity for each round is $\bigo{\frac{\threshold\log^2\threshold}{\confidence\approxerror^2\degree{}}}$ \degreeq{}, and $\bigo{\frac{\log(\log\threshold/\confidence)}{\approxerror^2}}$ \randedgeq{} queries. There are at most $\log^2\threshold$ such rounds, hence overall query bound for these values remain $\bigo{\frac{\threshold\log^4\threshold}{\confidence\approxerror^2\degree{}}}$ \degreeq{}, and $\bigo{\log^2\threshold\frac{\log(\log\threshold/\confidence)}{\approxerror^2}}$ \randedgeq{} queries. The bound on \randedgeq{} queries are independent of $\degbnd$, and hence remains $\bigo{\log^2\threshold\frac{\log(\log\threshold/\confidence)}{\approxerror^2}}$ if both the loops are executed entirely.


    Hence, we only consider the \degreeq{} queries for the lower values of $\degbnd$ such that $\degbnd \leq \nicefrac{\degree{}}{8}$. We argue that the probability of reaching any such value goes down as the value becomes lower. First observe that for any $\degbnd \leq \nicefrac{\degree{}}{8}$, the probability that it satisfies the valid $\degbnd$ condition (Line~\ref{Valid Degbnd}) is at least $1 - \confidence$ by union bound. Then, the probability to hit $\degbnd_i < \nicefrac{\degree{}}{2^j}$ is at most $\confidence^{j-2}$, as all the previous valid $\degbnd$ must have failed. Hence, we can bound the expected query complexity over the entire run of the algorithm as:
    
    \begin{align*}
        \confidence^{j-2}\cdot\frac{\constant\threshold\log^2\threshold}{2^j\confidence\approxerror^2\degree{}} = \frac{\constant\confidence^{j-3}\threshold\log^2\threshold}{2^j\approxerror^2\degree{}} = \bigo{\frac{\threshold\log^2\threshold}{\approxerror^2\degree{}}}
    \end{align*}

    Here, the last equality follows from the fact that $\confidence < \nicefrac{1}{2}$. Thus, the algorithm makes $\bigo{\frac{\threshold\log^4\threshold}{\confidence\approxerror^2\degree{}}}$ \degreeq{}, and $\bigo{\log^2\threshold\frac{\log(\log\threshold/\confidence)}{\approxerror^2}}$ \randedgeq{} queries in expectation.
\end{proof}

\subsection{Proofs for \texorpdfstring{\Cref{SubSec: Est w/o Advice}}{Section 5.3}}\label{AppSec: Proof for NoAdv}

\noadvlem*

\begin{proof}
    By~\Cref{Lem: DensEst Algo Guarantee}, the probability that at the $i$-th iteration, the check for good threshold (Line~\ref{ThresDetect}) succeeds for a threshold that is not good is at most $\nicefrac{\confidence}{2^{i+1}}$. Hence, by union bound, with probability at least $1 - \nicefrac{\confidence}{2}$, \thradv{} is called with a good threshold. Consequently, by~\Cref{Lem: Thr Advice Guarantee}, we have $\degest \in \fbrac{1\pm\approxerror}\degree{}$ with probability at least $1 - \confidence$.

    Now, we focus on the number of queries made by the algorithm. First, note that the algorithm makes $\bigot{\log(2^{i+2}/\confidence)}$ \degreeq{} and \randedgeq{} queries for each call of the \ldensest{}. The queries made up to the $i$-th iteration such that $\threshold_i \leq 8\arboricity$ can be bounded as $\bigot{\log(1/\confidence)}$. For the thresholds such that $\threshold > 8\arboricity$, observe that by~\Cref{Lemma: Arboricity Edge Bound} and \Cref{Lem: DensEst Algo Guarantee}, the probability that the algorithm reaches $\threshold_i > 2^{j+3}\arboricity$ is at most $\nicefrac{1}{2^{2j}}$, as each of the earlier good thresholds with $\ldens \geq \frac{3}{8}$ must have been missed for this to happen, and $\confidence \leq \nicefrac{1}{4}$. Then, we can bound the queries as $\bigot{\log(1/\confidence)}\sum_{j}2^{-j}j = \bigot{\log(1/\confidence)}$.
    

    
    The number of queries made by the call to \thradv{} depends on the value of $i$ for which the check for good threshold (Line~\ref{ThresDetect}) succeeds. If it is called for any threshold $\threshold \leq 8\arboricity$, it makes at most $\bigot{\frac{\arboricity}{\confidence\approxerror^2\degree{}}}$ \degreeq{} queries. 
    
    Now we consider the higher values, i.e., $\threshold > 8 \arboricity$. Again, we argue that the probability that the call is made with a high threshold decreases as the value increases. Recall that by~\Cref{Lem: Arboricity based Good Threshold}, and \ref{Lem: DensEst Algo Guarantee}, any threshold $\threshold \geq 8\arboricity$ passes the check for good threshold (Line~\ref{ThresDetect}) with probability at least $1 - \confidence$. Correspondingly, the probability that the threshold at which the check at Line~\ref{ThresDetect} succeeds is larger than $2^{j+4}\arboricity$ is at most $\nicefrac{1}{2^{j}}$ as the probability that each of the earlier checks failed is at most $\confidence < \nicefrac{1}{2}$. 
    
    Note that the only term in the query complexity of \thradv{} is $\threshold\log^3\threshold$. By \Cref{lem:expected-stopping-cost}, we can bound its expectation as $\bigot{\arboricity}$. Similarly, we can also bound the number of \randedgeq{} queries made for the \thradv{} call as $\bigot{\frac{\log(1/\confidence)}{\approxerror^2}}$. Hence, the expected query complexity for the \thradv{} call is $\bigot{\frac{\arboricity}{\confidence\approxerror^2\degree{}}}$ \degreeq{} queries, and $\bigot{\frac{\log(1/\confidence)}{\approxerror^2}}$ \randedgeq{} queries, by~\Cref{Lem: Thr Advice Guarantee}.
\end{proof}

\begin{lemma}[Expected Stopping-Cost bound of \noadv{}]
\label{lem:expected-stopping-cost}
Suppose the stopping threshold $\tau$ satisfies
$\Pr[\tau = 2^{j+4}\arboricity] \le 2^{-2j}$ for all integers $j\ge 0$.
Then we have $\mathbb{E}[\tau\log^{4}\tau] = O(\arboricity\log^{4}\arboricity)$. In particular, $\mathbb{E}[\tau\log^{4}\tau]=\tilde{O}(\arboricity)$.
\end{lemma}

\begin{proof}
First we use the upper bound on the probability to obtain:
\begin{align*}
\mathbb{E}[\tau\log^{4}\tau]
&= \sum_{j\ge 0} \Pr[\tau=2^{j+4}\arboricity]\cdot 2^{j+4}\arboricity
     (\log(2^{j+4}\arboricity))^{4} 
\leq \constant\arboricity \sum_{j\ge 0} 2^{-j}\, (\log\arboricity + j)^{4}.
\end{align*}

We analyse the sum in two parts, for $j\leq \lceil\log_{2}\arboricity\rceil$ and $j > \lceil\log_{2}\arboricity\rceil$. We denote $L = \lceil\log_{2}\arboricity\rceil$.

For the lower range, i.e. $0\le j\le L$, we have $\log\arboricity + j = O(\log\arboricity)$, hence
\begin{align*}
\sum_{j=0}^{L}2^{-j}(\log\arboricity + j)^{4}
= O(\log^{4}\arboricity)\sum_{j=0}^{L}2^{-j}
= O(\log^{4}\arboricity).
\end{align*}

For the higher range, i.e. $j>L$, we have $(\log\arboricity + j)^{4} = O(j^{4})$. In this case, we have:
\[
\sum_{j=L+1}^{\infty}2^{-j}(\log\arboricity + j)^{4}
= O\!\left(\sum_{j=0}^{\infty}2^{-j} j^{4}\right) = O(1),
\]
where the final sum is a finite constant~\cite{GKP1994Concrete}. Combining the two parts gives
\[
\sum_{j\ge 0}2^{-j}(\log\arboricity + j)^{4}
= O(\log^{4}\arboricity).
\]
Therefore we have $\mathbb{E}[\tau\log^{4}\tau]
\le \arboricity \cdot O(\log^{4}\arboricity)
= O(\arboricity\log^{4}\arboricity)$, as claimed.
\end{proof}

\subsection{Proofs for \texorpdfstring{\Cref{SubSec: Final}}{Section 5.4}}\label{AppSec: Proof for Final}

Next, we state a result due to Beretta and T\v{e}tek~\cite{BerettaTetek/TALG/2024/BetterSumEstimationViaWeightedSampling} that can estimate average degree when at most a constant fraction of vertices are isolated. For exposition purpose, we state the result under a slightly more stricter assumption that no vertex is isolated.

\begin{lemma}[Edge Estimation~\cite{BerettaTetek/TALG/2024/BetterSumEstimationViaWeightedSampling,BerettaCS/ArXiv/2025/FasterEdgeEstimation}]\label{Lem: Beretta Tetek Algo}
    There exists an algorithm that, given access to \degreeq{} and \randedgeq{} queries to a graph $\graph$ with no isolated vertices, and average degree $\degree{}$, and an advice $\degub \geq \degree{}$ outputs an estimate $\degest$ such that $\degest \in (1\pm\approxerror)\degree{}$ of the average degree $\degree{}$ with probability at least $\frac{9}{10}$ if the advice is correct. If $\degub \leq \frac{\degree{}}{2^i}$, the algorithm rejects the advice with probability at least $1 - \frac{1}{2^{i-1}}$. The algorithm makes $\bigot{\frac{\degub}{\approxerror^2}}$ queries in expectation.
\end{lemma}

We can combine this result with our \noadv{} algorithm. We use the algorithm of~\cite{BerettaTetek/TALG/2024/BetterSumEstimationViaWeightedSampling,BerettaCS/ArXiv/2025/FasterEdgeEstimation} with $\degub = \sqrt{\threshold}$ at each guess of the threshold, and output the average degree if the algorithm returns an estimate $\degest$, or move to the next threshold if the advice is rejected. The other modification concerns the call to the \thradv{} algorithm once a good threshold $\threshold^*$ has been detected. In this case, we check only up to $\degbnd \geq \sqrt{\threshold^*}$, and otherwise invoke the algorithm of~\cite{BerettaTetek/TALG/2024/BetterSumEstimationViaWeightedSampling,BerettaCS/ArXiv/2025/FasterEdgeEstimation} with $\degub = \sqrt{\threshold^*}$. These modifications yield the following result, established through an argument similar to that of~\Cref{Lemma: No Advice Guarantee}.

\UBNI*

\section{Proofs for \texorpdfstring{\Cref{Sec: Lower Bounds}}{Section 6}}\label{AppSec: Proof for Lower bounds}

\subsection{Proofs for \texorpdfstring{\Cref{SubSec: GeneralLB}}{Section 6.1}}\label{AppSec: Generic LB Proofs}

The proofs are divided in two parts. In~\Cref{Thm: Lower Bound - High Degree}, we show that $\bigomega{\frac{\arboricity}{\degree{}}}$ queries are necessary for the high degree case, i.e. when $\degree{} \geq \sqrt{\arboricity}$. Then, in~\Cref{Thm: Lower Bound - Low Degree}, we show that $\bigomega{\degree{}}$ queries are necessary for the low degree case, i.e. when $\degree{} \leq \sqrt{\arboricity}$. We start with the higher degree case~\Cref{Thm: Lower Bound - High Degree}.

\begin{figure*}[ht!]

    \centering

\begin{tikzpicture}[
  vertex/.style={circle, draw, fill=blue!15, inner sep=1.5pt},
  matchingnode/.style={circle, draw, fill=red!15, inner sep=2pt},
  matchingedge/.style={thick, red},
  cliqueedge/.style={thin, gray!70},
  scale=1
]

\newcommand{\drawclique}[1]{
  \foreach \i in {1,...,5}
    \node[vertex] (v\i#1) at ($(#1)+(72*\i:0.7)$) {};
  \foreach \i in {1,...,5}{
    \foreach \j in {\i,...,5}{
      \ifnum\i<\j
        \draw[cliqueedge] (v\i#1) -- (v\j#1);
      \fi
    }
  }
}

\coordinate (C1) at (0,0);
\coordinate (C2) at (0,-2);
\coordinate (C3) at (0,-5); 

\drawclique{C1}
\drawclique{C2}
\drawclique{C3}

\draw[dotted, thick] (0,-3) -- (0,-4.2);

\draw[decorate,decoration={brace,amplitude=8pt}] 
  (-1.6,-6) -- (-1.6,0.8);
\node[rotate=90, anchor=south] at (-1.9,-2.6)
  {$\lbclqcnt$ cliques of size $\arboricity$};

\begin{scope}[xshift=5cm]

  \node[matchingnode] (m1a) at (0,0) {};
  \node[matchingnode] (m1b) at (1,0) {};
  \draw[matchingedge] (m1a) -- (m1b);

  \node[matchingnode] (m2a) at (0,-1.25) {};
  \node[matchingnode] (m2b) at (1,-1.25) {};
  \draw[matchingedge] (m2a) -- (m2b);

  \node[matchingnode] (m3a) at (0,-2.5) {};
  \node[matchingnode] (m3b) at (1,-2.5) {};
  \draw[matchingedge] (m3a) -- (m3b);

  \draw[dotted, thick] (0.5,-3.4) -- (0.5,-4.2);

  \node[matchingnode] (m4a) at (0,-5) {};
  \node[matchingnode] (m4b) at (1,-5) {};
  \draw[matchingedge] (m4a) -- (m4b);

  \draw[decorate,decoration={brace,amplitude=8pt}]
    (2.5,0.8) -- (2.5,-6);
  \node[rotate=90, anchor=north] at (2.8,-2.6)
        {matching on $\fbrac{\vertexcount - \lbclqcnt\arboricity}$ vertices};

\end{scope}

\end{tikzpicture}
    
    \caption*{The construction for \Cref{Thm: Lower Bound - High Degree,Thm: Lower Bound - Low Degree,Thm: Lower Bound})}
\end{figure*}

\begin{restatable}{theorem}{lbhighdeg}[Lower Bound - High Degree Case]\label{Thm: Lower Bound - High Degree}
    Consider $\vertexcount$,  $\arboricity$, and $\degree{}$ such that $\arboricity/4 \geq \degree{} \geq \sqrt{\arboricity}$ and $\degree{} \geq 4$, and there exists a graph with $\vertexcount$ vertices, average degree $\degree{}$, and arboricity $\arboricity$. Then, there exists a graph with $\vertexcount$ vertices, average degree $\degree{}$, arboricity $\arboricity$, and no isolated vertices such that any algorithm that can obtain a $(1\pm\approxerror)$-multiplicative approximation of the average degree $\degree{}$ of the graph with probability at least $\frac{3}{4}$ for any $\approxerror < \frac{1}{3}$:
    
    \begin{enumerate}
        \item Using \degreeq{}, \neighbourq{}, and \randedgeq{} queries must make at least $\bigomega{\frac{\arboricity}{\degree{}}}$ queries.
        \item Using \degreeq{}, \neighbourq{}, \edgeexistsq{} and \randedgeq{} queries must make at least $\bigomega{\frac{\arboricity}{\degree{}}}$ queries, given $\arboricity \leq \sqrt{\vertexcount}$.
        \item Using \degreeq{}, \neighbourq{}, \edgeexistsq{}, \fullnbrq{} and \randedgeq{} queries must make at least $\bigomega{\frac{\arboricity}{\degree{}}}$ queries, given $\arboricity \leq \vertexcount^{2/5}$.
    \end{enumerate}
\end{restatable}

\begin{proof}
    We first describe the graph that we use to construct the lower bound. The graphs consist of $\lbclqcnt \geq 1$ cliques of size $\arboricity$, and the remaining vertices form a matching. We denote the set of vertices forming the clique (resp. matching) to be $\lbclqv$ (resp. $\lbmatv$). Similarly, we denote the set of edges forming the clique (resp. matching) to be $\lbclqe$ (resp. $\lbmate$). Note that by our construction, we have $\size{\lbclqv} = \lbclqcnt\arboricity$, $\size{\lbmatv} = \vertexcount - \lbclqcnt\arboricity$, $\size{\lbclqe} = \bigtheta{\lbclqcnt\arboricity^2}$, and $\size{\lbmate} = \bigtheta{\vertexcount - \lbclqcnt\arboricity}$. We denote the set of vertices obtained by making \degreeq{} queries as $\lbdegq$, and the endpoints of the edges obtained by making \randedgeq{} queries as $\lbreq$.

    For the lower bound instances, we consider two cases, one where $\lbclqcnt = \frac{\vertexcount\degree{}}{\arboricity^2}$, and the other where $\lbclqcnt = \frac{2\vertexcount\degree{}}{\arboricity^2}$. Note that this must be larger than $1$ by~\Cref{Lemma: Arboricity Edge Bound}. Hence, $\lbclqcnt \geq 1$ is satisfied. Furthermore, note that $\lbclqcnt \leq \frac{\vertexcount}{8}$ as we have $\degree{} \leq \frac{\arboricity}{4}$, and $\arboricity \geq \degree{} \geq 4$. Note that any algorithm that obtains a $(1\pm\approxerror)$-multiplicative approximation for an $\approxerror \leq \frac{1}{3}$ to the average degree $\degree{}$ should be able to distinguish between these two cases. 

    Now, we want to show that across these two cases, the samples received are indistinguishable if the algorithm makes $\lbq = \smallo{\frac{\arboricity}{\degree{}}}$ queries. We first show that in this case, $\lbdegq \subseteq \lbmatv$, $\lbreq \subseteq \lbclqv$, and all these samples are distinct with probability $1 - \smallo{1}$. Next, we show that all the samples obtained are distinct, i.e. there are no collisions. Hence, the two cases remain indistinguishable.

    First, we show that $\lbdegq \subseteq \lbmatv$ with probability $1 - \smallo{1}$. The probability that a \degreeq{} query hits a vertex in $\lbclqv$ is $\frac{\size{\lbclqv}}{\vertexcount} = \frac{\lbclqcnt\arboricity}{\vertexcount} \leq \frac{2\degree{}}{\arboricity}$, as $\lbclqcnt \leq \frac{2\vertexcount\degree{}}{\arboricity^2}$. Hence, the probability that any of the $\lbq$ \degreeq{} queries hits a vertex in $\lbclqv$ is $\frac{2\lbq\degree{}}{\arboricity} = \smallo{1}$. Furthermore, all the samples in $\lbdegq$ are distinct with probability $1 - \smallo{1}$ as $\sqrt{\lbmatv} = \sqrt{\vertexcount - \frac{2\vertexcount\degree{}}{\arboricity}} \geq \sqrt{\frac{\vertexcount}{2}} \geq \frac{\arboricity}{2\sqrt{\degree{}}} \geq \frac{\arboricity}{\degree{}}$. Here, the first inequality follows from the fact that $\degree{} \leq \frac{\arboricity}{4}$, the second inequality follows from the fact that $\arboricity \leq \sqrt{\vertexcount\degree{}}$, and the last inequality follows from the fact that $\degree{} \geq 4$.

    Next, we show that $\lbreq \subseteq \lbclqv$ with probability $1 - \smallo{1}$. The probability that a \randedgeq{} query hits an edge in $\lbmate$ is $\frac{\size{\lbmate}}{\vertexcount\degree{}} \leq \frac{\vertexcount - \vertexcount\degree{}/\arboricity}{\vertexcount\degree{}} \leq \frac{1}{\degree{}}$. Hence, the probability that any of the $\lbq$ \randedgeq{} queries hits an edge in $\lbmate$ is at most $\frac{\lbq}{\degree{}} = \smallo{1}$, as $\degree{} \geq \sqrt{\arboricity}$. Furthermore, all the endpoints in $\lbreq$ are distinct with probability $1 - \smallo{1}$ as $\sqrt{\size{\lbclqv}} = \sqrt{\lbclqcnt\arboricity} = \sqrt{\frac{\vertexcount\degree{}}{\arboricity}} = \frac{\sqrt{\vertexcount\degree{}}}{\sqrt{\arboricity}} \geq \sqrt{\arboricity} \geq \frac{\arboricity}{\degree{}}$. Here, the second last inequality follows from~\Cref{Lemma: Arboricity Edge Bound}, and the last inequality follows from the fact that $\degree{} \geq \sqrt{\arboricity}$,.

    Now that we have shown that the output of the \degreeq{} and \randedgeq{} queries are indistinguishable between the two cases considered, we now prove the three parts of the result by showing that the \neighbourq{}, \edgeexistsq{}, and \fullnbrq{} queries indistinguishable within the given ranges of $\arboricity$.
    
    \noindent\textit{Proof of part (1): }We start with the \neighbourq{} queries. First we consider the \neighbourq{} queries on the vertices in $\lbmatv$. We assume that the algorithm does not make redundant \neighbourq{} queries, i.e. does not make \neighbourq{} queries on the same vertex twice. In that case, as $\sqrt{\size{\lbmatv}} \geq \frac{\arboricity}{\degree{}}$, there is no collision in $\lbq$ queries with probability $1 - \smallo{1}$.

    Next, we consider the \neighbourq{} queries on the vertices in $\lbclqv$. Again, we assume that the algorithm does not make redundant \neighbourq{} queries, i.e. does not make \neighbourq{} queries on the same vertex twice. In that case, even if there is only a single clique, it has $\arboricity$ vertices. Note that as $\degree{} \geq \sqrt{\arboricity}$, , we have $\sqrt{\arboricity} \geq \frac{\arboricity}{\degree{}}$. Hence, there will not be a collision in $\lbq$ queries with probability $1 - \smallo{1}$. Therefore, the algorithm can not distinguish the two cases using $\smallo{\frac{\arboricity}{\degree{}}}$ \degreeq{}, \neighbourq{}, and \randedgeq{} queries. This completes the proof of part (1) of the claim.

    \noindent\textit{Proof of part (2): }Next, we consider the \edgeexistsq{} queries. We show that if the algorithm makes $\lbq$ \edgeexistsq{} queries, then all such queries return $0$ with probability $1 - \smallo{1}$ for both instances. The algorithm can not gain any information using \edgeexistsq{} on vertices in $\lbmatv$ beyond that obtained through \neighbourq{} queries. Hence, we consider the algorithm does not make these redundant queries, and the \edgeexistsq{} queries are made only on vertices in $\lbclqv$. Let us consider the queries to be $\sequence{e}{\lbq}$ where each $e_i = (\altvertex_i,\vertex_i)$ for some $\altvertex_i, \vertex_i \in \lbclqv$. The probability that any such query returns $1$ is at most $\frac{1}{\lbclqcnt}$. Hence, in $\lbq$ queries, the expected number of queries to return $1$ is $\frac{2\lbq}{\lbclqcnt} \leq \smallo{\frac{\arboricity^3}{\vertexcount\degree{}^2}} \leq \smallo{\frac{\arboricity^2}{\vertexcount}} = \smallo{1}$. Here, the last inequality follows from the fact that $\degree{} \geq \sqrt{\arboricity}$, and the last follows from the fact that $\arboricity = \smallo{\sqrt{\vertexcount}}$. Then, by Markov's Inequality, the no queries returns $1$ with probability $1 - \smallo{1}$. This completes the proof for part (2).

    \noindent\textit{Proof of part (3): }Next, we consider the \fullnbrq{} queries. Again for any vertex in $\lbmatv$, \fullnbrq{} is equivalent to \neighbourq{}, and we consider the algorithm does not make such redundant queries. For vertices in $\lbclqv$, each \fullnbrq{} query reveal all the id-s in the clique containing the vertex. There are $\lbclqcnt = \frac{\vertexcount\degree{}}{\arboricity^2}$ cliques, and  we have $\sqrt{\frac{\vertexcount\degree{}}{\arboricity^2}} \geq \sqrt{\arboricity^{1/2}\degree{}} \geq \sqrt{\arboricity^{1/2}\degree{}\frac{\arboricity^{3/2}}{\degree{}^3}} = \frac{\arboricity}{\degree{}}$. Here, the first inequality follows from the fact that $\vertexcount \geq \arboricity^{5/2}$, and the second inequality follows from the fact that $\degree{} \geq \sqrt{\arboricity}$. Hence, with probability $1-\smallo{1}$, the  $\lbq = \smallo{\frac{\arboricity}{\degree{}}}$ \fullnbrq{} queries does not hit any clique twice. Hence, the all the cliques obtained by \fullnbrq{} queries are distinct, and the two cases remain indistinguishable. This completes the proof for part (3).
\end{proof}

Next, we state our lower bound for the low degree case.

\begin{restatable}{theorem}{lblowdeg}[Lower Bound - Low Degree Case]\label{Thm: Lower Bound - Low Degree}
    Consider $\vertexcount$,  $\arboricity$, and $\degree{}$ such that $\degree{} \leq \sqrt{\arboricity} \leq \vertexcount^{1/3}$ and $\degree{} \geq 4$, and there exists a graph with $\vertexcount$ vertices, average degree $\degree{}$, and arboricity $\arboricity$. Then, there exists a graph with $\vertexcount$ vertices, average degree $\degree{}$, arboricity $\arboricity$, and no isolated vertices such that any algorithm that can obtain a $(1\pm\approxerror)$-multiplicative approximation of the average degree $\degree{}$ of the graph with probability at least $\frac{3}{4}$ for any $\approxerror < \frac{1}{3}$:
    
        \begin{enumerate}
        \item Using \degreeq{}, \neighbourq{}, and \randedgeq{} queries must make at least $\bigomega{\degree{}}$ queries.
        \item Using \degreeq{}, \neighbourq{}, \edgeexistsq{} and \randedgeq{} queries must make at least $\bigomega{\degree{}}$ queries, given $\arboricity \leq \sqrt{\vertexcount}$.
        \item Using \degreeq{}, \neighbourq{}, \edgeexistsq{}, \fullnbrq{} and \randedgeq{} queries must make at least $\bigomega{\degree{}}$ queries, given $\arboricity \leq \vertexcount^{2/5}$.
    \end{enumerate}
\end{restatable}

\begin{proof}
    The proof uses the same construction as~\Cref{Thm: Lower Bound - High Degree}, we consider the graphs to contain $\lbclqcnt$ cliques, and the remaining vertices form a matching. The two instances considered has $\lbclqcnt$ set to $\frac{\vertexcount\degree{}}{\arboricity^2}$ and $\frac{2\vertexcount\degree{}}{\arboricity^2}$. This value is still valid as $1 \leq \lbclqcnt \leq \frac{\vertexcount}{8}$ in this parameter regime as well. In this case, for contradiction, we assume that there exists an algorithm that makes $\lbq = \smallo{\degree{}}$ queries. As before, we show that any such algorithm can not distinguish between the two cases.

    First, we consider the case of $\degreeq{}$ queries. The probability that any of them hit a vertex in $\lbclqv$ is at most $\frac{2\degree{}}{\arboricity}$. The probability that any of the $\lbq$ \degreeq{} queries hit $\lbclqv$ is at most $\frac{2\lbq\degree{}}{\arboricity}$, which is $\smallo{1}$, as $\degree{} \leq \sqrt{\arboricity}$. As before, note that $\sqrt{\size{\lbmatv}} \geq \frac{\arboricity}{\degree{}}$. This, given we consider $\degree{} \leq \sqrt{\arboricity}$, ensures that all the samples are distinct with probability $1 - \smallo{1}$.

    Next, we consider \randedgeq{} queries, the probability that any such query hits an edge in $\lbmate$ is, as before, at most $\frac{1}{\degree{}}$. Hence, the probability that any of the $\lbq$ \randedgeq{} queries hits an edge in $\lbmate$ is $1 - \smallo{1}$. We now consider whether all the endpoints of $\lbreq$ are distinct. Here, we have $\sqrt{\size{\lbclqv}} = \sqrt{\frac{\vertexcount\degree{}}{\arboricity}} \geq \sqrt{\frac{\arboricity^{3/2}\degree{}}{\arboricity}} = \sqrt{\arboricity\degree{}} \geq \degree{}$. Here, the first inequality follows from the fact that $\arboricity \leq \vertexcount^{2/3}$, and the second inequality follows from the fact that $\degree{} \leq \sqrt{\arboricity}$. Now we focus on the other queries.

    \noindent\textit{Proof of part (1): }Now, we consider the \neighbourq{} queries. As before, we concern ourselves only with the $\neighbourq{}$ queries on the vertices in $\lbclqv$. Even if we have only a single clique, as $\sqrt{\arboricity} \geq \degree{}$, there is no collision with probability $1 - \smallo{1}$. Thus, with probability $1-\smallo{1}$, any algorithm that makes $\lbq = \smallo{\degree{}}$ queries can not distinguish between the two cases.
    
    \noindent\textit{Proof of part (2): }Next, we consider the \edgeexistsq{} queries. Again, the \edgeexistsq{} queries to vertices in $\lbmatv$ does not yield any additional information compared to \neighbourq{} queries. The probability that any \edgeexistsq{} queries on the vertex pairs of $\lbclqv$ returns $1$ is at most $\frac{\lbq}{\lbclqcnt} = \smallo{\frac{\arboricity^2}{\vertexcount}} = \smallo{1}$, where the last equality follows from the fact that $\arboricity \leq \sqrt{\vertexcount}$.

    \noindent\textit{Proof of part (2): }We now consider the \fullnbrq{} queries. As before, we only need to consider the queries made on the vertices in $\lbclqv$. To ensure that there is no collision with probability $1 - \smallo{1}$, we must have $\sqrt{\lbclqcnt} \geq \degree{}$. This holds true as $\sqrt{\frac{\vertexcount\degree{}}{\arboricity^2}} \geq \sqrt{\arboricity^{1/2}\degree{}} \geq \degree{}$. Here, the first inequality is due to the fact that $\arboricity \leq \vertexcount^{2/5}$, and the second inequality is due to the fact that $\degree{} \leq \sqrt{\arboricity}$.
\end{proof}

Combining~\Cref{Thm: Lower Bound - High Degree}~and~\ref{Thm: Lower Bound - Low Degree}, we obtain the main lower bound result of our work. The key observation here is the fact that $\degree{} \geq \nicefrac{\arboricity^2}{\vertexcount}$ (\Cref{Lemma: Arboricity Edge Bound}) ensures that for $\arboricity \geq \vertexcount^{2/3}$, we always have $\degree{} \geq \nicefrac{\arboricity}{\degree{}}$.

\LB*

\subsection{Proofs for \texorpdfstring{\Cref{SubSec: EpsDelLB}}{Section 6.2}}\label{AppLBSec: EpsDel Proof}

We now introduce the popcount thresholding problem \cite{DBLP:conf/approx/AssadiN22} that we use for our proofs. 

\begin{definition}[$\defptp$]\label{Def: PTP}
    Given a string $\alicestring \in \sbrac{0,1}^\stringlength$, the problem $\defptp$ is to  distinguish whether $\alicestring$ is  generated by i.i.d. samples from $\ptpdone$ or $\ptpdtwo$ defined as follows:
    \begin{itemize}
        \item \textbf{$\ptpdone$ :} For all $i \in \tbrac{\stringlength}$, $\alicestring_i$ is set to $1$ with probability $\fbrac{1-2\ptpsep}\frac{\ptpprob}{\stringlength}$, and set to $0$, otherwise.
        \item \textbf{$\ptpdtwo$ :} For all $i \in \tbrac{\stringlength}$, $\alicestring_i$ is set to $1$ with probability $\fbrac{1+2\ptpsep}\frac{\ptpprob}{\stringlength}$, and set to $0$, otherwise.
    \end{itemize}
    The problem is to decide if $\alicestring$ is generated from $\ptpdone$ or $\ptpdtwo$ by querying $\alicestring$ at any of its $\stringlength$ bits.
\end{definition}

We now state two results regarding the \ptp{} problem \cite{DBLP:conf/approx/AssadiN22,bishnu2025arboricityrandomedgequeries}.

\begin{lemma}[\ptp{} Query Lower Bound]\label{Lemma: PTP Query Lower Bound}
    For any $\ptpsep \in (0,1/4)$, $\confidence \in (0,1/100)$, and integers $\stringlength \geq 1$, $\log(1/\confidence)\cdot12/\ptpsep^2\leq \ptpprob \leq \stringlength/6$, any randomized algorithm that solves the $\defptp$ problem must make at least $\frac{\stringlength\log\fbrac{1/4\confidence}}{24\ptpsep^2\ptpprob}$ in the worst-case.
\end{lemma}

\begin{restatable}{lemma}{numones}\label{Lemma: PTP Deviation Conditional Bound}
     In $\defptp$, for any $\ptpsep \in (0,1/4)$, $\confidence \in (0,1/100)$, and integers $\stringlength \geq 1$, $\log(1/\confidence)\cdot12/\ptpsep^2\leq \ptpprob \leq \stringlength/6$, let $\norm{\alicestring}_1$ denote the number of $1$'s in the string $\alicestring$, then:
   \begin{enumerate}
    \item[(i)] $\Pr\left[\|\alicestring\|_1 > (1 - \ptpsep) \cdot \ptpprob \,\middle|\, \alicestring \sim \ptpdone\right] \leq \confidence$;
    \item[(ii)] $\Pr\left[\|\alicestring\|_1 < (1 + \ptpsep) \cdot \ptpprob \,\middle|\, \alicestring \sim \ptpdtwo\right] \leq \confidence$;
    \item[(iii)] $\Pr\left[\|\alicestring\|_1 < (1 - 4\ptpsep) \cdot \ptpprob \right] \leq \confidence$; \end{enumerate}   
\end{restatable}

Now, we establish our $\approxerror,\confidence$-sensitive lower bound. The construction of the graph is outlined in \Cref{Fig: LBEpsDel}.

\begin{figure*}[ht]
    \centering

\begin{tikzpicture}[
  vertex/.style={circle, draw, fill=blue!15, inner sep=1.5pt},
  cliqueedge/.style={thin, gray!70},
  scale=1
]

\newcommand{\drawclique}[1]{
  \foreach \i in {1,...,5}
    \node[vertex] (v\i#1) at ($(#1)+(72*\i:0.7)$) {};
  \foreach \i in {1,...,5}{
    \foreach \j in {\i,...,5}{
      \ifnum\i<\j
        \draw[cliqueedge] (v\i#1) -- (v\j#1);
      \fi
    }
  }
}

\newcommand{\drawindep}[1]{
  \foreach \i in {1,...,5}
    \node[vertex] (v\i#1) at ($(#1)+(72*\i:0.7)$) {};
}


\node at (1.5, 2.5) {\Large $x_1 = 1$};
\node at (4.5, 2.5) {\Large $x_2 = 0$};
\node at (7.5, 2.5) {\Large $x_3 = 1$};
\node at (10, 2.5) {\Large $\dots$};
\node at (12.5, 2.5) {\Large $x_\stringlength = 0$};

\draw[dashed, thick] (1.5,2.15) -- (1.5,1.35);
\draw[dashed, thick] (4.5,2.15) -- (4.5,1.35);
\draw[dashed, thick] (7.5,2.15) -- (7.5,1.35);
\draw[dashed, thick] (12.5,2.15) -- (12.5,1.35);

\coordinate (V1) at (1.5, 0);
\coordinate (V2) at (4.5, 0);
\coordinate (V3) at (7.5, 0);
\coordinate (Vs) at (12.5, 0);

\drawclique{V1}
\drawindep{V2}
\drawclique{V3}
\drawindep{Vs}

\node at (1.5, -1.5) {$V_1$};
\node at (4.5, -1.5) {$V_2$};
\node at (7.5, -1.5) {$V_3$};
\node at (12.5, -1.5) {$V_\stringlength$};

\node[text width=2.5cm, align=center] at (1.5, -2.6) {Clique\\(size $\alpha$)};
\node[text width=2.5cm, align=center] at (4.5, -2.6) {Independent Set\\(size $\alpha$)};
\node[text width=2.5cm, align=center] at (7.5, -2.6) {Clique\\(size $\alpha$)};
\node[text width=2.5cm, align=center] at (12.5, -2.6) {Independent Set\\(size $\alpha$)};

\draw[dashed, gray] (0.2, 1.2) rectangle (2.8, -1.8);
\draw[dashed, gray] (3.2, 1.2) rectangle (5.8, -1.8);
\draw[dashed, gray] (6.2, 1.2) rectangle (8.8, -1.8);
\draw[dashed, gray] (11.2, 1.2) rectangle (13.8, -1.8);

\end{tikzpicture}

\caption*{The construction for \Cref{Theorem: Lower Bound Epsilon Delta}, mapping the \ptp{} string $x \in \sbrac{0,1}^\stringlength$ to the graph $G_x$.}
\end{figure*}

\LBEpsDel*

\begin{proof}
    We proof by contradiction, let us assume there exists an algorithm $\sA_{\degree{}}$ that uses $\smallo{\frac{1}{\approxerror^2}}$ \randedgeq{} query and $\smallo{\frac{\arboricity}{\approxerror^2\degree{}}\log\fbrac{\frac{1}{\confidence}}}$ \degreeq{}, \neighbourq{}, \edgeexistsq{}, and \fullnbrq{} queries that can obtain a $(1\pm\approxerror)$-approximation of the average degree $\degree{}$ of the graph with probability at least $1 - \confidence$.

    We now show that such an algorithm can be used to solve the $\defptp{}$ problem using less than $\frac{\stringlength\log\fbrac{3/4\confidence}}{24\ptpsep^2\ptpprob}$ queries, for $\ptpsep \leq \frac{1}{8}$, and $\log(3/\confidence)\cdot12/\ptpsep^2\leq \ptpprob \leq \stringlength/6$. We associate each string $\alicestring$ with a graph $\graph_\alicestring$ of $\stringlength\arboricity$ vertices. The vertex set $\vertexset_\alicestring$ can be partitioned into $\stringlength$ parts $\sequence{\vertexset}{\stringlength}$, each of size $\arboricity$. For each element $\alicestring_i \in \alicestring$, the set of vertices $\vertexset_i$ forms a clique if $\alicestring_i = 1$, and is an independent set otherwise. By this construction, the number of edges of $\graph_\alicestring$ is $\binom{\arboricity}{2}\norm{\alicestring}_1$, and arboricity is $\arboricity$. By \Cref{Lemma: PTP Deviation Conditional Bound}, we have that the average degree is $\bigomega{\frac{\ptpprob\arboricity}{\stringlength}}$. Hence, by \Cref{Lemma: PTP Deviation Conditional Bound}, for $\log(3/\confidence)\cdot12/\ptpsep^2\leq \ptpprob \leq \stringlength/6$ any algorithm that can obtain an $\fbrac{1\pm\ptpsep}$-approximation of the average degree with probability at least $1 - \frac{\confidence}{3}$ should be able to solve the $\defptp{}$ problem with probability $1 - \confidence$.

    To establish the contradiction, we need to show that the algorithm $\sA_{\degree{}}$ can be simulated using $\frac{\stringlength\log(3/4\confidence)}{200\ptpsep^2\ptpprob}$ queries to the \ptp{} string $\alicestring$. We describe how each of the queries available to $\sA_{\degree{}}$ can be simulated using queries to $\alicestring$. We begin with describing the \degreeq{}, \neighbourq{}, \edgeexistsq{} and \fullnbrq{} queries.

    \begin{itemize}
        \item \textbf{\degreeq{}:} For a vertex $\vertex \in \vertexset_i$, return $\arboricity-1$ if $\alicestring_i = 1$, and $0$ otherwise. Random vertex can be chosen by choosing a number randomly from $[\stringlength\arboricity]$.
    
        \item \textbf{\neighbourq{}:} For a vertex $\vertex \in \vertexset_i$, return the $j$-th vertex in $\vertexset_i/\sbrac{\vertex}$ if $\alicestring_i = 1$, and $0$ otherwise.
    
        \item \textbf{\edgeexistsq{}:} If the vertices are in different $\vertexset_i,\vertexset_j$, return $0$. If both are in same $\vertexset_i$, return $1$ if $\alicestring_i = 1$, and $0$ otherwise.
    
        \item \textbf{\fullnbrq{}:} For a vertex $\vertex \in \vertexset_i$, return $\vertexset_i/\sbrac{\vertex}$ if $\alicestring_i = 1$, and return $\emptyset$ otherwise.
    \end{itemize}

    Hence, each of these queries can be simulated $1$ query to $\alicestring$. Then, the $\smallo{\frac{\arboricity}{\approxerror^2\degree{}}\log\fbrac{\frac{1}{\confidence}}}$ \degreeq{}, \neighbourq{}, \edgeexistsq{}, and \fullnbrq{} queries required for $\sA_{\degree{}}$ can be simulated using $\frac{\stringlength\log(3/4\confidence)}{200\ptpsep^2\ptpprob}$ queries to $\alicestring$. 
    
    Next, we describe how to simulate \randedgeq{} queries. To generate an edge uniformly at random, we first sample a random clique, and then sample a random edge from that clique. To sample the random clique, we check random indices of $\alicestring$ till we find a $1$. The probability of a random index $i$ having $\alicestring_i = 1$ is:
    \[
    \frac{\norm{\alicestring}_1}{\stringlength} \geq \frac{(1-4\ptpsep)\ptpprob}{\stringlength} \geq \frac{\ptpprob}{2\stringlength}
    \]
    Here, the first inequality follows from \Cref{Lemma: PTP Deviation Conditional Bound}, and the second inequality follows from the fact that $\ptpsep \leq \frac{1}{8}$. By Chernoff bound, with probability $1 - \frac{\confidence}{3}$, $\smallo{\frac{1}{\approxerror^2}}$ \randedgeq{} queries can be simulated using $\frac{\stringlength\log(1/4\confidence)}{200\ptpsep^2\ptpprob}$ queries to $\alicestring$. Hence, by an union bound argument, if there exists an algorithm that uses $\smallo{\frac{1}{\approxerror^2}}$ \randedgeq{} query and $\smallo{\frac{\arboricity}{\approxerror^2\degree{}}\log\fbrac{\frac{1}{\confidence}}}$ \degreeq{}, \neighbourq{}, \edgeexistsq{}, and \fullnbrq{} queries that can obtain a $(1\pm\approxerror)$-approximation of the average degree $\degree{}$ of the graph with probability at least $1 - \confidence$, then there exists an algorithm that uses at most $\frac{\stringlength\log(3/4\confidence)}{200\ptpsep^2\ptpprob}$ queries to solve $\defptp$ with probability at least $1 - \confidence$. Hence, by \Cref{Lemma: PTP Query Lower Bound}, a contradiction.
\end{proof}

\section{Concentration Inequalities}\label{App: ConcIneq}

Here, we state some well-known concentration inequalities that is relevant to our work. These bounds are widely known~\cite{Mitzenmacher_Upfal_2005,DubhashiP/Book/2009/RandAlgInequalities}.

\begin{lemma}[Markov's Inequality]\label{Lemma: Markov's Inequality}
    For a non-negative random variable $X$, we have:
    \begin{align*}
        \Pr\tbrac{X \geq t} \leq \frac{\E[X]}{t}
    \end{align*}
\end{lemma}

\begin{lemma}[Chebyshev's Inequality]\label{Lemma: Chebyshev's Inequality}
    For a  random variable $X$, we have:
    \begin{align*}
        \Pr\tbrac{\abs{X -\E[X]} \geq \approxerror} \leq \frac{\Var[X]}{t^2}
    \end{align*}
\end{lemma}

\begin{lemma}[Multiplicative Chernoff Bound]\label{Lemma: Multiplicative Chernoff Bound}
    Given i.i.d. random variables $X_1,X_2,...,X_t$ where $\Pr[X_i = 1] = p$ and $\Pr[X_i = 0] = (1-p)$, define $X = \sum_{i \in [t]} X_i$. Then, we have:
    \begin{align*}
    \Pr[\abs{X-\E[X]} \geq \approxerror\E\tbrac{X}] &\leq 2\exp{\fbrac{-\frac{\approxerror^2\E\tbrac{X}}{2}}} & 0 \leq \approxerror <1
    \end{align*}
\end{lemma}

\begin{lemma}[Additive Chernoff Bound]\label{Lemma: Additive Chernoff Bound}
    Given i.i.d. random variables $X_1,X_2,...,X_t$ where $\Pr[X_i$ $ = 1] = p$ and $\Pr[X_i = 0] = (1-p)$, define $X = \sum_{i \in [t]} X_i$. Then, we have:
    \begin{align*}
    \Pr[\abs{X-\E[X]} \geq \approxerror t] &\leq 2\exp{\fbrac{-\frac{\approxerror^2t}{2}}} & 0 \leq \approxerror
    \end{align*}
\end{lemma}



\section{Extending to Approximate \randedgeq{} Query}\label{App: SimEdge}

In this section, we show that \Cref{Thm: Main UB - General} can be extended to approximate random edge queries. Note that our algorithm only uses \randedgeq{} for calls to \ldensest{}. Hence, it suffices to show that \ldensest{} can be extended for such queries. First, we formally define approximate random edge queries:

\begin{itemize}
    \item{\nrandedgeq{}:} This query returns an edge $\edge \in \edgeset$ from a distribution $\sD'_\edgeset$ such that for every edge $(\altvertex,\vertex)$:
    \begin{align*}
        \Pr_{\edge \sim \sD'_\edgeset}\tbrac{\edge = (\altvertex,\vertex)} \in \frac{(1\pm\edgenoise)}{\edgecount}
    \end{align*}
\end{itemize}

We present the algorithm \nldensest{} that takes as input a possible threshold $\threshold$, the edge sample noise parameter $\edgenoise$, and the number of queries $\coinq$ it makes, and access to \degreeq{} and \nrandedgeq{} queries, and outputs $\estldens$ as an estimate of the quantity $\ldens$.

\begin{algorithm}
    \caption{\nldensest{$\fbrac{\threshold,\qcolor{\coinq}}$}}\label{Alg: Light Density Estimator Noisy}
    \begin{algorithmic}[1]
        \Require Access to \degreeq{} and \nrandedgeq{} queries, a threshold $\threshold$, and  number of queries $\coinq$
        \For{$i \in [\qcolor{\coinq}]$}
            \State $(\altvertex,\vertex) \gets \nrandedgeq{}$ \Comment{Makes \qcolor{$\coinq$} \nrandedgeq{} queries}
            \State $w \gets \uniform\sbrac{\altvertex,\vertex}$
            \State $X_i \gets$ $1$ if $w \in \lightverts$ (i.e. $\degreeq{\fbrac{w}} \leq \threshold$), and $0$ otherwise(i.e. $\degreeq{\fbrac{w}} > \threshold$) \Comment{Makes \qcolor{$\coinq$} \degreeq{} queries}
        \EndFor
        \State \Return  $\estldens \gets \frac{1}{\coinq} \sum_{i\in[\coinq]} X_i$
    \end{algorithmic}
\end{algorithm}

The next lemma provides the performance guarantees of \nldensest{}. Here, the performance of the algorithm depends on the edge sample noise parameter $\edgenoise$ as well as the number of queries made and the quantity $\ldens$.

\begin{lemma}\label{Lem: DensEst Algo Guarantee Noisy}
    For the algorithm \nldensest{}, the following holds:
    \begin{enumerate}
        \item $\E\tbrac{\estldens} \in \ldens\fbrac{1\pm\edgenoise}$
        \item If $\coinq = \bigomega{\log\fbrac{1/\confidence}}$, $\edgenoise \leq \frac{1}{64}$, and $\ldens < \frac{1}{4}$, i.e., the threshold $\threshold$ is not good, $\estldens \leq \frac{5}{16}$ with probability at least $1 - \confidence$.
        \item If $\coinq = \bigomega{\log\fbrac{1/\confidence}}$, $\edgenoise \leq \frac{1}{64}$, and $\ldens \geq \frac{3}{8}$, the estimate satisfies $\estldens \geq \frac{5}{16}$ with probability at least $1 - \confidence$.
        \item If the threshold $\threshold$ is good, $\edgenoise \leq \frac{\approxerror}{2}$, and $\coinq = \bigomega{\frac{\log\fbrac{1/\confidence}}{\fbrac{\approxerror-\edgenoise}^2}}$, $\estldens \in (1\pm\approxerror)\ldens$ with probability at least $1 - \confidence$.
    \end{enumerate}
\end{lemma}

\begin{proof}
    First, we note that the $X_i$-s are defined in~\nldensest{} are i.i.d. Bernoulli random variables with $p \in \ldens (1\pm\edgenoise)$.
    \begin{align*}
        \Pr[X_i = 1] \in \sum_{\lightvert \in \lightverts} \sum_{i \in [\degree{\lightvert}]} \frac{\fbrac{1\pm\edgenoise}}{2\edgecount} \in \fbrac{1\pm\edgenoise}\sum_{\lightvert \in \lightverts} \frac{\degree{\lightvert}}{2\edgecount} \in \frac{\ledgecnt(1\pm\edgenoise)}{2\edgecount} \in \ldens(1\pm\edgenoise)
    \end{align*}
    The result of part (1) follows directly by taking expectation. We obtain the parts (2) and (3) of the claim through an additive Chernoff bound with additive error $\nicefrac{1}{64}$. For the part (4), as the threshold is good, we have $\ldens \geq \nicefrac{1}{4}$, and obtaining an $(1 \pm \frac{\approxerror}{4})$ approximation to $p$ suffices to obtain an $(1 \pm \approxerror)$ approximation to $\ldens$. Hence, by multiplicative Chernoff bound, the claim holds.
\end{proof}

We now state the following lemma on approximate edge sampling due to Eden et al.~\cite{EdenRR/ICALP/2019/ArboricitySamplingEdges}.

\begin{lemma}\label{Lem: Approx Edge Sampling}
    Given the number of vertices $\vertexcount$, and arboricity $\arboricity$ of a graph $\graph$, \nrandedgeq{} can be realised using $\bigo{\frac{\arboricity}{\edgenoise\degree{}}\log^3\vertexcount}$ \degreeq{}, and \neighbourq{} queries.
\end{lemma}

Combining \Cref{Lem: DensEst Algo Guarantee Noisy,Lem: Approx Edge Sampling,Thm: Main UB - General}, we obtain the following result.

\begin{theorem}\label{Thm: Arboricity UB}
    There exists a randomized algorithm that, the number of vertices $\vertexcount$, and arboricity $\arboricity$ of a graph $\graph$ with average degree $\degree{}$, outputs an  estimate $\degest$ such that $\degest \in (1\pm\approxerror)\degree{}$ with probability at least $\frac{9}{10}$ using $\bigo{\frac{\arboricity}{\approxerror^3\degree{}}\poly{\log\vertexcount}}$ \degreeq{} and \neighbourq{} queries.
\end{theorem}

Next, we consider the following result on exact edge sampling due to Eden et al.~\cite{eden2022samplingedgesublineartime}.

\begin{lemma}\label{Lem: Exact Edge Sampling}
    Given the number of vertices $\vertexcount$ of a graph $\graph$, \randedgeq{} can be realised using $\bigo{\sqrt{\frac{\vertexcount}{\degree{}}}}$ \degreeq{}, and \neighbourq{} queries.
\end{lemma}

Combining \Cref{Lem: Exact Edge Sampling,Thm: Main UB - General}, and the fact that $\arboricity = \bigo{\sqrt{\edgecount}}$ we obtain the following result.

\begin{theorem}\label{Thm: Gen UB}
    There exists a randomized algorithm that, the number of vertices $\vertexcount$ of a graph $\graph$ with average degree $\degree{}$, outputs an  estimate $\degest$ such that $\degest \in (1\pm\approxerror)\degree{}$ with probability at least $\frac{9}{10}$ using $\bigo{\frac{1}{\approxerror^2}\sqrt{\frac{\vertexcount}{\degree{}}}\poly{\log\vertexcount}}$ \degreeq{} and \neighbourq{} queries.
\end{theorem}

\end{document}